\documentclass[12pt]{article}
\pdfoutput=1
\usepackage{jheppub} 
\usepackage{tikz-cd}
\usepackage{graphicx,amsmath,amssymb,amsthm,multirow,array,bm,esint,epsfig,hyperref}
\usetikzlibrary{decorations.markings}
\tikzset{middlearrow/.style={
        decoration={markings,
            mark= at position 0.5 with {\arrow{#1}} ,
        },
        postaction={decorate}
    }
}

 \title{Island Paradigm and Information Recovery from Radiation}

\author{Krishna Jalan,}
\author{Roji Pius,}
\author{Manish Ramchander}

\affiliation[]{The Institute of Mathematical Sciences, IV Cross Road, C.I.T. Campus, Taramani, Chennai, India 600113}

 \affiliation[]{Homi Bhabha National Institute, Training School Complex, Anushakti Nagar, Mumbai, India 400094}

\emailAdd{krishnajalan@imsc.res.in}
\emailAdd{rojipius@imsc.res.in}
\emailAdd{manishd@imsc.res.in}

\abstract{ 
The island paradigm for an AdS$_2$ eternal black hole in the Hartle-Hawking state coupled to a finite temperature non-gravitating bath asserts that after the Page time the operators in the black hole interior can be reconstructed using the bath degrees of freedom. We demonstrate  this assertion by consideing a special operator $\mathbb{U}_{bath}$ that has nontrivial action only on the bath degrees of freedom. We show via the gravitational Euclidean path integral analysis that though before the Page time  $\mathbb{U}_{bath}$ does not translate the operators in the bath to the black hole interior, after the Page time it takes them to the black hole interior.  }

\begin{document} 
\maketitle
\flushbottom

 
\section{Introduction}
\label{sec:intro}

One of the most significant developments towards a consistent resolution of black hole information paradox \cite {Hawking:1976ra, Hawking:1975vcx, Page:1993df} is the island paradigm, which posits that the black hole evaporation is a unitary evolution, and it is possible to reconstruct all the information that went into the black hole using the Hawking radiation collected at the future null infinity of the spacetime \cite{Almheiri:2019hni, Almheiri:2020cfm}. The simplest setup where the island paradigm has been studied is of \cite{Almheiri:2019yqk}, of an AdS$_2$ eternal black hole in thermal equilibrium with a non-gravitating bath, where it makes two assertions.  First is that before the Page time the operators in the interior of the black hole can be reconstructed using the operators in the left and right exterior of the black hole within the gravitational region. Second assertion is that,  after the Page time, the interior of the black hole cannot be reconstructed this way; the reconstruction requires  the operators in the non-gravitating bath  \cite{Almheiri:2019yqk}.


 The island paradigm can be demonstrated if one  identifies two operators $\mathbb{A}(u)$ and $\mathbb{B}(u)$, where $u$ is the boundary time,  having the following properties in the semiclassical limit\footnote{In \cite{Penington:2019kki} the second assertion of the island paradigm is already demonstrated using the Petz map.}.  The operators operators $\mathbb{A}(u)$ and $\mathbb{B}(u)$ act only the degrees of freedom localised outside the black hole horizon but within  the gravitational region and on the degrees of freedom associated with the bath respectively. In addition, they have the property that before the Page time the action of $\mathbb{A}(u)$ transports the gravitationally dressed local operators in the gravitational region outside the horizon to the black hole interior and the action $\mathbb{B}(u)$ keeps the local operators in the bath region there itself. On the contrary after the Page time the action of $\mathbb{B}(u)$ transports the local operators inside the bath region to the black hole interior and the action of $\mathbb{A}(u)$  keeps the gravitationally dressed local operators in the gravitational region  itself.  
  
In this paper, we substantiate the second assertion of the island paradigm by constructing an example for the operator $\mathbb{B}(u)$ in the simplest set up of AdS$_2$ eternal black hole in thermal equilibrium with the  non-gravitating baths attached to the right and left boundaries of the black hole spacetime. We  denote this operator  as $\mathbb{U}_{bath}(t,u)$. Since the black hole spacetime is in thermal equilibrium with a bath, there is a constant exchange of degrees of freedom across their interface. Consequently, an arbitrary local operator which is declared as a bath degree of freedom at time $u_1$ may not be a bath degree of freedom at a later time $u_2$. This implies that an operator that acts only on bath degrees of freedom has to be defined with an explicit dependence on time $u$.  

We  define the operator $\mathbb{U}_{bath}(t,u)$ with an explicit time dependence as follows
 \begin{equation}
 \label{Ubath}
 \mathbb{U}_{bath}(t,u)= \bm{\rho}^{\text{i}t}_{{M_u^{bath}}'}\bm{\rho}_{{M}_u^{bath}}^{-\text{i}t}\bm{\rho}^{-\text{i}t}_{{{N}_{u}^{bath}}'}\bm{\rho}_{N^{bath}_{u-a}} ^{\text{i}t}  \qquad\forall ~t\in \mathbb{R}.
 \end{equation}
 The operator  $\bm{\rho}_{{M_u^{bath}}}$ can be understood as  the reduced density matrix in  full quantum JT gravity coupled to matter associated with  an interval  $M_u^{bath}=\left(q_M,i^R_0\right)$ that lies within the bath region in the slice $B_u$. The slice $B_u$ is the union of equal time slices $B^L_u$ and $B_u^R$ in the left and right baths. The point $q_M$ is at a finite location in $B_u^R$ and the point $i^R_0$ is the spatial infinity of  $B_u^R$. The operator  $\bm{\rho}_{{M_u^{bath}}}$ is defined via the appropriate analytic continuation of the gravitational Euclidean path integral, which involves sum over all asymptotically Euclidean AdS$_2$  hyperbolic geometries  that are attached to  a Euclidean bath  with a cut along the interval $M_0^{bath}$. Similarly,  the operators $\bm{\rho}_{{M_u^{bath}}'}$, $\bm{\rho}_{{N_{u-a}^{bath}}}$, $\bm{\rho}_{{N_{u}^{bath}}'}$  are the reduced density matrices  associated with the intervals ${M_u^{bath}}'$, $N_{u-a}^{bath}$, ${N_{u}^{bath}}'$ respectively. The interval ${M_u^{bath}}'$ is  complement to the interval $M_u^{bath}$ in the slice $B_u$ that is restricted to the bath region without any intersection with the gravitational region.  The interval  $N_{u-a}^{bath}=\left(q_N,i^R_0\right)$ lies in another slice $B_{u-a}$, where $a$ is a positive real number.  the The  interval ${N_{u}^{bath}}'$  is same as the  interval ${M_{u}^{bath}}'$. The intervals   $M_u^{bath}$ and  $N_{u-a}^{bath}$ are chosen such that their end points $q_M$ and $q_N$ are connected by a light like geodesic\footnote{The form of  $\mathbb{U}_{bath}(t,u)$ is inspired by the interior reconstruction proposal in \cite{Leutheusser:2021qhd, Leutheusser:2021frk} based on the idea of half-sided translation \cite{ Borchers:2000pv}.}.
 
    We show that this operator can't translate operators in the bath to the black hole interior before the Page time.  Furthermore, using the gravitational path integral we argue that the action of $\mathbb{U}_{bath}(t,u)$ on a local operator in bath changes after the Page time due to a change in the saddle of the gravitational path integral. By choosing a specific conformal matter theory we demonstrate that the wormhole saddle enables $\mathbb{U}_{bath}(t,u)$ to spread the operators in the bath to the black hole interior.  Hence, $\mathbb{U}_{bath}(t,u)$ can be considered as an example for the operator $\mathbb{B}(u)$.

The organisation of the paper is as follows. Section \ref{sect 2} contains a brief discussion of the island construction in the context of an AdS$_2$ eternal black hole in thermal equilibrium with a non-gravitating finite temperature bath.  In section \ref{sect 22} we discuss the implications of the island paradigm and the role of the operator $ \mathbb{U}_{bath}(t,u)$ in recovering the black hole information from the bath radiation in the semiclassical limit. We delineate the JT-gravity path integral description for $ \mathbb{U}_{bath}(t,u)$ in section \ref{sect 3}. In section \ref{sect 4} we  show that before the Page time  the operator  $ \mathbb{U}_{bath}(t,u)$ in the semiclassical limit fails to  translate operators in the bath to the black hole interior and demonstrate that the free energy of the gravitational saddle grows linearly with respect to the boundary time. Moreover, we construct a wormhole saddle having time independent free energy after the Page time. In section \ref{sect 5}, we explain how this wormhole saddle  enables $ \mathbb{U}_{bath}(t,u)$  to spread an operator localised  in the bath to the blackhole interior. We conclude in section \ref{sect 6} by summarising the result and mention an interesting future direction that deserves further investigation. 


\section{Island paradigm for a black hole in equilibrium with a bath}\label{sect 2}


  \begin{figure}
\centering
\begin{tikzpicture}[scale=1.15]
    \draw [red,  thick] plot [smooth] coordinates {(-6,4.75) (-3.5,3.5) (-1,4.75)} ;
    \draw [thick,](1 mm, 10 pt) (-3.5,7)--(1.5,2.5)node[pos=.5,sloped, above] {$w^+=\infty,~ y_R^+=\infty$} ;
        \draw [thick,](1 mm, 10 pt) (-3.5,-2)--(1.5,2.5) node[pos=.5,sloped, below] {$w^-=-\infty, ~ y_R^-=-\infty$}  ;
      \draw  [red,  thick] plot[smooth]coordinates {(-6,0.25) (-3.5,1.5) (-1,0.25)} ;
     \draw [thick,](1 mm, 10 pt) (-3.5,7)--(-8.5,2.5) node[pos=.5,sloped, above] {$y_L^-=\infty,~w^-=\infty$} ;
        \draw [thick,](1 mm, 10 pt) (-3.5,-2)--(-8.5,2.5) node[pos=.5,sloped, below] {$y_L^+=-\infty, ~w^+=-\infty$}; 
\draw (1.75,2.5) node {$i_0^R$};
\draw (-8.75,2.5) node {$i_0^L$};
  \draw [very thick](1 mm, 10 pt) (-6,4.75)--(-1,.25) node[pos=.425,sloped, below] {$w^+=0, ~y_L^+=\infty\qquad \quad ~ y_R^+=-\infty$};
 \draw [very thick](1 mm, 10 pt) (-6,.25)--(-1,4.75)node[pos=.55,sloped, below] {$y_L^-=-\infty \qquad w^-=0,~y_R^-=\infty$};
 \draw[thick](1 mm, 10 pt) (-6,.25)--(-6,4.75) node[pos=.5,sloped, above] {$w^+w^-=-e^{\frac{4\pi \epsilon}{\beta}}$};
  \draw[thick](1 mm, 10 pt) (-1,.25)--(-1,4.75)node[pos=.5,sloped, below] {$w^+w^-=-e^{\frac{4\pi \epsilon}{\beta}}$};
\end{tikzpicture}
\caption{An eternal black hole in equilibrium with a finite temperature bath can be described using a plane with lightcone coordinates $(w^+,w^-)$. The right/left Rindler wedge describes the right/left side of the black exterior coupled to the right/left bath having flat metric. The right/left Rindler wedge can be covered using the lightcone coordinates $(y_{R/L}^+,y_{R/L}^-)$. The lines $w^+w^-=0$ are the future and the past horizons of the eternal black hole. The interface between the black hole and the bath satisfy the equation $w^+w^-=-e^{\frac{4\pi \epsilon}{\beta}}$. The red  hyperbola represents the singularity on which the dilaton profile vanishes. \label{fig:BHAdS2bath}} 
\end{figure}

 Consider  an AdS$_2$ eternal black hole solution in JT gravity with inverse temperature $\beta$ glued to a non-gravitating bath having the same inverse temperature.  The  matter in the spacetime is chosen to be a CFT having central charge $c$. To make the computations manageable, we assume that the matter CFT is not directly coupled to the dilaton. The thermal equilibrium guarantees that the classical geometry of the black hole spacetime is the same as that of an isolated eternal black hole.  The glued system can be conveniently described as a region in a plane with the Kruskal  coordinates \footnote{The AdS Schwarzschild coordinate for the right and left exteriors of the black hole   $(y_R^+,y_R^-)$ and $(y_L^+,y_L^-)$ respectively. They also cover the left and the right bath.  The relation between the two coordinates are given by $ w^{\pm}=\pm e^{\pm \frac{2\pi y_{R}^{\pm}}{\beta}}, \qquad  w^{\pm}=\mp e^{\mp \frac{2\pi y_{L}^{\pm}}{\beta}}$. } $(w^+,w^-)$ \cite{Almheiri:2019yqk}, see figure \ref{fig:BHAdS2bath}. The AdS$_2$ black hole region (the gravitational region) in the $w$-plane is given by the equation 
\begin{equation} 
w^+w^-\geq -e^{\frac{4\pi \epsilon}{\beta}}
\end{equation}
 where $\epsilon$ is a real parameter that specifies the location of the cut-out boundary of AdS$_2$. The bath is the remaining region in the $w$-plane.  The  black hole horizon is given by the equation $$w^+w^-=0$$ and the past and the future singularities of the black hole lie on the hyperbolas on which the following dilaton profile vanishes
 \begin{equation}\label{dilaton}
\phi(w^+,w^-)=\phi_0+\frac{2\pi \phi_r}{\beta}\frac{1-w^+w^-}{1+w^+w^-}
\end{equation}
where the first term $\phi_0$ give rises to the extremal entropy, and $\frac{\phi_r}{\epsilon}$ is the boundary value of the difference $\phi(w^+,w^-)-\phi_0$.  In this coordinate system, the metric in the AdS$_2$ region is given by 
\begin{equation} 
ds^2_{BH}=\frac{4dw^+dw^-}{(1+w^+w^-)^2}
\end{equation}
and the metric in the bath region is given by 
\begin{equation} 
ds^2_{B}=\frac{\beta^2}{4\pi^2\epsilon^2}\frac{dw^+dw^-}{w^+w^-}.
\end{equation}
 Note that the geometry of an AdS$_2$ eternal black hole in equilibrium with a thermal bath  is the same as that of an AdS$_2$ eternal black hole with reflecting boundary conditions at the boundary. However, quantum mechanically the two systems have drastic differences.  The constant exchange of matter between the black hole and the bath gives rise to an ever-growing entanglement between them.  In fact, a  short computation of the bath entropy using the replica method \cite{Calabrese:2004eu} reveals that at late times the entanglement entropy of both the black hole and the bath rises linearly with respect to the boundary time $u$  \cite{Almheiri:2019yqk}
\begin{align}\label{bathentropy}
S_{\text{blackhole}}(u)=S_{\text{bath}}(u)\xrightarrow{u\gg \beta} \frac{2\pi c }{3\beta} u.
\end{align}
The linear growth  makes the von-Neumann entanglement entropy bigger than the thermal entropy of the black hole at sufficiently large times.\par

The paradoxical growth of the von Neumann entropy can be resolved by changing the saddle of the gravitational replica path integral from the trivial saddle to the wormhole saddle having minimum free energy. After the Page time  the saddle having the least free energy is a Euclidean wormhole geometry \cite{Almheiri:2019yqk, Almheiri:2019qdq}.  The replica wormhole introduces a non-trivial entanglement wedge island, containing the black hole interior,  for the bath. The inclusion of this island into the entropy computation of the bath converts the linear growth of bath entropy to a constant, equal to the thermal entropy of the black hole. The physical interpretation of this phenomenon is that at late times, the black hole interior can be completely reconstructed using the left and right bath. 


\section{The operator \texorpdfstring{$\mathbb{U}_{bath}(t,u)$}{Lg} } \label{sect 22}
The island paradigm makes two assertions in the simplest set up of AdS$_2$ eternal black hole in thermal equilibrium with the  non-gravitating baths. The first one is that after the Page time the operators in the interior of  an AdS$_2$ eternal black hole in equilibrium with a finite temperature non-gravitating bath can not be reconstructed using the operators in the black hole region outside the  horizon. The second assertion is that after the Page time the operators in the black hole interior can be reconstructed using the bath degrees of freedom.  Since the black hole spacetime is  in thermal equilibrium with the bath, there is a constant exchange of degrees of freedom between the bath and the black hole.  Consequently, an arbitrary local operator localised within bath/black hole region at time $u_1$ may not stay within the same region at a later time $u_2$. At each time $u$ one must check the region of support of an operator before declaring it as a local operator within  bath/black hole.  This implies that an operator that acts only on bath/black hole degrees of freedom has to be defined with an explicit dependence on time $u$.  Therefore the assertions of the island paradigm can be demonstrated if one  identifies two operators  $\mathbb{A}(u)$ and $\mathbb{B}(u)$ in quantum JT gravity coupled to matter with explicit time dependence,  having the following properties in the semiclassical limit
  \begin{itemize}
 \item Before the Page time,  denoted as $u_{Page}$, $\mathbb{A}(u)$  acts only the degrees of freedom in  the left and right exterior of the black hole within the gravitational region and $\mathbb{B}(u)$ acts only the degrees of freedom within the bath. 
 \item  Assume that $\mathcal{O}_{bh}(u)$ is a dressed local operator in the exterior of the black hole within the gravitational region.  Then the transformed operator $$\mathbb{A}^{-1}(u)\mathcal{O}_{bh}(u)\mathbb{A}(u)$$ for $u<u_{Page}$, must be an operator localised in the black hole interior. However,  it  must be an operator localised  in the exterior of the black hole within the gravitational region after the Page time.
 \item If $\mathcal{O}_{bath}(u)$ is an operator localised in the bath, then $$\mathbb{B}^{-1}(u)\mathcal{O}_{bath}(u)\mathbb{B}(u)$$ must be an operator localised in the bath  for $u<u_{Page}$.  However, for $u<u_{Page}$,  it must be an operator localised in the interior of the black hole, the island region.
 \end{itemize}


\begin{figure}
\centering
\begin{tikzpicture}[scale=1]
    \draw [red,  thick] plot [smooth] coordinates {(-6,4.75) (-3.5,3.5) (-1,4.75)} ;
    \draw [thick,](1 mm, 10 pt) (-3.5,7)--(1.5,2.5) ;
        \draw [thick,](1 mm, 10 pt) (-3.5,-2)--(1.5,2.5) ;
      \draw  [red,  thick] plot[smooth]coordinates {(-6,0.25) (-3.5,1.5) (-1,0.25)} ;
     \draw [thick,](1 mm, 10 pt) (-3.5,7)--(-8.5,2.5) ;
        \draw [thick,](1 mm, 10 pt) (-3.5,-2)--(-8.5,2.5) ; 
   \fill[fill=white!20!gray] (-8.5,2.5)--(-7,1.15)--(-6,2.05)--(-7.45,3.45);
     \fill[fill=white!20!gray]  (-1,2.05)--(0,1.15)--(1.5,2.5)--(.5,3.45);
      \draw [blue, very thick,dashed] plot [smooth] coordinates {(-8.5,2.5) (-5.5,2) (-3.5,2.5)(-1.5,2) (1.5,2.5)} ;
        \draw [violet, very thick,dashed] plot [smooth] coordinates {(-8.5,2.5) (-5.5,1.35) (-3.5,2.5)(-1.5,1.35) (1.5,2.5)} ;
\draw (.5,2) node {$\Diamond_{{B}^R_u}$};
\draw (-7,2) node {$\Diamond_{{B}_u^L}$};
\draw (1.75,2.5) node {$i_0^R$};
\draw (-8.75,2.5) node {$i_0^L$};
  \draw [very thick](1 mm, 10 pt) (-6,4.75)--(-1,.25);
 \draw [very thick](1 mm, 10 pt) (-6,.25)--(-1,4.75);
 \draw[thick](1 mm, 10 pt) (-6,.25)--(-6,4.75) node[pos=.5,sloped, below] {};
  \draw[thick](1 mm, 10 pt) (-1,.25)--(-1,4.75)node[pos=.5,sloped, above] {};
  \draw (-1.35,2.1) node {$q_M$};
  \draw (-6.3,2.05) node[above] {$b$};
    \draw (-4.5,2.25) node[above] {$\Sigma_u$};
      \draw (-5.25,1.35) node[above] {$\Sigma_{u-a}$};
  \draw (-.5,1.6) node[below] {$q_N$};
  \draw (-1,2.05) node  [circle,fill,inner sep=1pt]{};
  \draw (-.5,1.6) node  [circle,fill,inner sep=1pt]{};
  \draw (-6,2.05) node  [circle,fill,inner sep=1pt]{};
\end{tikzpicture}
\caption{ The causal diamonds  $\Diamond_{B^L_u}$ and $\Diamond_{B^R_u}$ are the domains of dependence of the  largest intervals $B_u^L$ and $B_u^R$  on the slice $\Sigma_u$ that lie entirely within the left and the right  bath respectively. The points $q_M$ and $q_N$ of the intervals $M_u^{bath}$ and $N_{u-a}^{bath}$ lie in the slices $\Sigma_u$ and $\Sigma_{u-a}$ respectively.  \label{fig:bathalgebra}}
\end{figure}

In this section, we will  describe an operator that behaves as $\mathbb{B}(u)$  in the above setup of AdS$_2$ black hole in equilibrium with a bath\footnote{ The first assertion was demonstrated  in \cite{Jalan:2023dmq} by following this interpretation of the island paradigm.}. Assume that $\Sigma_u$ is a Cauchy slice   parametrised by the AdS$_2$ boundary time $u$. It is given by the equation 
\begin{equation}\label{sliceeq} 
\frac{w^+}{w^-}=\begin{cases} -e^{\frac{4\pi u}{\beta}} & \text{on the right patch} \\ -e^{-\frac{4\pi u}{\beta}} & \text{on the left patch} \end{cases}.
\end{equation}
 Let  $B_u^L$ and $B_u^R$ be the  largest intervals  on the slice $\Sigma_u$ that lie entirely within the left and the right  bath respectively.   We denote the domains of dependence of $B_u^L$ and $B_u^R$ as the causal diamonds  $\Diamond_{B^L_u}$ and $\Diamond_{B^R_u}$, see figure \ref{fig:bathalgebra}.  Assume that $M_u^{bath}$ is the interval $(q_M,i_0^R)$ in $B_u^R$.  The interval within $B_u=B_u^L\cup B_u^R$ that is complement to $M_u^{bath}$ is the interval ${M^{bath}_u}'=(i_0^L,b)$.  Similarly, consider  the interval $N_{u-a}=(q_N,i_0^R)$ in  $B_{u-a}$, where $a$ real number. The complementary interval $\left(i_o^L,b\right)$ within $B_{u}$ is denoted as ${N^{bath}_{u}}'$. The domain of dependence of the intervals $M_u^{bath}, {M^{bath}_u}', N^{bath}_{u-a}$ and ${N^{bath}_{u}}'$ are the causal diamonds   $\Diamond_{M_u^{bath}}, \Diamond_{{M^{bath}_u}'},\Diamond_{N^{bath}_{u-a}}$ and $\Diamond_{{N^{bath}_{u}}'}$ respectively.  The coordinates of the points $q_M,q_N$ and $b$ are
 \begin{align}
 (w^+_{q_M}, w^-_{q_M}) &= \left(e^{\frac{2\pi }{\beta}(u-\epsilon)},-e^{-\frac{2\pi }{\beta}(u+\epsilon)}\right)\nonumber\\
 (w^+_{q_N}, w^-_{q_N}) &= \left(e^{\frac{2\pi }{\beta}(u-\epsilon)},-e^{-\frac{2\pi }{\beta}(u-2a+\epsilon)}\right)\nonumber\\
  (w^+_{b}, w^-_{b}) &= \left(-e^{-\frac{2\pi }{\beta}(u+\epsilon)},e^{\frac{2\pi }{\beta}(u-\epsilon)}\right)\nonumber\\
 \end{align}


 \begin{figure}
\centering
\begin{tikzpicture}[scale=1]
    \draw [red,  thick] plot [smooth] coordinates {(-6,4.75) (-3.5,3.5) (-1,4.75)} ;
    \draw [thick,](1 mm, 10 pt) (-3.5,7)--(1.5,2.5) ;
        \draw [thick,](1 mm, 10 pt) (-3.5,-2)--(1.5,2.5) ;
      \draw  [red,  thick] plot[smooth]coordinates {(-6,0.25) (-3.5,1.5) (-1,0.25)} ;
     \draw [thick,](1 mm, 10 pt) (-3.5,7)--(-8.5,2.5)  ;
        \draw [thick,](1 mm, 10 pt) (-3.5,-2)--(-8.5,2.5) ; 
   \fill[fill=black!50!green] (-1,2.05)--(0,1.15)--(1.5,2.5)--(.5,3.45);
   \fill[fill=blue!20] (-.5,1.6)--(0,1.15)--(1.5,2.5)--(1,2.95);
      \fill[fill=black!10!yellow] (-8.5,2.5)--(-7,1.15)--(-6,2.05)--(-7.45,3.45);
\draw (.8,2.5) node[above] {$\Diamond_{{M}^{bath}_u}$};
\draw (.8,1.5) node[above] {$\Diamond_{{N}^{bath}_{u-a}}$};
\draw (-7,2.5) node[above] {$\Diamond_{{{M}'}^{bath}_u}=\Diamond_{{{N}'}_{u}^{bath}}$};
  \draw (-1.35,2.1) node {$q_M$};
  \draw (-5.7,2.1) node {$b$};
    \draw (-4.5,2.25) node[above] {$\Sigma_u$};
  \draw (-.5,1.6) node[below] {$q_N$};
  \draw [very thick](1 mm, 10 pt) (-6,4.75)--(-1,.25) ;
 \draw [very thick](1 mm, 10 pt) (-6,.25)--(-1,4.75);
 \draw[thick](1 mm, 10 pt) (-6,.25)--(-6,4.75);
  \draw[thick](1 mm, 10 pt) (-1,.25)--(-1,4.75);
  \draw (1.75,2.5) node {$i_0^R$};
\draw (-8.75,2.5) node {$i_0^L$};
        \draw [blue, very thick,dashed] plot [smooth] coordinates {(-8.5,2.5) (-5.5,2) (-3.5,2.5)(-1.5,2) (1.5,2.5)} ;
\end{tikzpicture}
\caption{ The  operator $\mathbb{U}_{bath}(t,u)$ is defined using the reduced density matrices in the full quantum theory associated with an arbitrary time  slice in the causal diamonds $\Diamond_{{M}^{bath}_u}, \Diamond_{{N}^{bath}_{u-a}}, \Diamond_{{{M}'}^{bath}_u}$ and $\Diamond_{{{N}'}_{u}^{bath}}$. \label{fig:BHAdS2bathMN1}}
\end{figure}

We propose the following operator as a candidate for $\mathbb{B}(u)$  
 \begin{align}
 \label{Ubath1}
 \mathbb{U}_{bath}(t,u)= \bm{\rho}^{\text{i}t}_{{M_u^{bath}}'}\bm{\rho}_{{M}_u^{bath}}^{-\text{i}t}\bm{\rho}^{-\text{i}t}_{{{N}_{u}^{bath}}'}\bm{\rho}_{N^{bath}_{u-a}} ^{\text{i}t}  \qquad\forall ~t\in \mathbb{R}
 \end{align}
 where  the operator $\bm{\rho}_{A}$ can be formally understood as the reduced density matrix in the full quantum theory associated with a cut along the interval  $A$.   We define $\bm{\rho}_{A}$ via the quantum Euclidean JT path integral. The JT gravity path integral for the operator $\bm{\rho}_{I_0^{bath}}$ at time $u=0$ involves sum over all asymptotically Euclidean AdS$_2$  hyperbolic geometries  that are attached to  a Euclidean bath  with a cut along the interval $I_0^{bath}$, where $I_u^{bath}=M_u^{bath},{M_u^{bath}}',{N_u^{bath}}'$. The $\bm{\rho}_{I_u^{bath}}$ for nonzero vales of $u$ is obtained by  the analytic continuation of Euclidean time  to $\text{i}u$ .    Similarly, $\bm{\rho}_{N_{u-a}^{bath}}$   is obtained by the analytic continuation of the Euclidean time  to $\text{i}(u-a)$. In the $G_N\to 0$ limit, before the Page time,  these density matrices are given by the gravitational path integrals on the euclidean $w$-planes with cuts in the bath region as shown in figures \ref{fig:Dpathintegral} and \ref{fig:Dpathintegral2}.

   \begin{figure}
\centering
		\begin{tikzpicture}[scale=2]
			\fill[draw=black,fill=cyan!5] (-3,2) rectangle (-5,0);
			\fill[draw=black, fill=teal!30] (-4,1) circle[radius=0.5];
			\draw[blue, very thick] (-3,1) -- (-3.5,1);
			\draw (-2.5,1.2) node {$\rho_{{M_0^{bath}}}$};
			\fill[draw=black,fill=cyan!5] (-5.5,2) rectangle (-7.5,0);
			\fill[draw=black, fill=teal!30] (-6.5,1) circle[radius=0.5];
			\draw[cyan,very thick] (-7,1) --(-7.5,1);
			\draw (-8,1.2) node {$\rho_{{M_0^{bath}}'}$};
	\draw (-7.1,1.2) node {$b$};
			\draw (-3.3,1.2) node {$q_M$};
					\end{tikzpicture}
\caption{The path integral representation of $\rho_{{M_0^{bath}}}$ and   $\rho_{{M_0^{bath}}'}$ which are the semiclassical limit of $\bm{\rho}_{{M_0^{bath}}}$ and   $\bm{\rho}_{{M_0^{bath}}'}$ respectively.} \label{fig:Dpathintegral}
\end{figure}


   \begin{figure}
\centering
		\begin{tikzpicture}[scale=2]
			\fill[draw=black,fill=cyan!5] (-3,2) rectangle (-5,0);
			\fill[draw=black, fill=teal!30] (-4,1) circle[radius=0.5];
			\draw[green, very thick] (-3,1) -- (-3.3,1);
			\draw (-2.5,1.2) node {$\rho_{{N_u^{bath}}}$};

			\fill[draw=black,fill=cyan!5] (-5.5,2) rectangle (-7.5,0);
			\fill[draw=black, fill=teal!30] (-6.5,1) circle[radius=0.5];
			\draw[brown,very thick] (-7,1) --(-7.5,1);
			\draw (-8,1.2) node {$\rho_{{N_u^{bath}}'}$};
			\draw (-7.1,1.2) node {$b$};
			\draw (-3.3,1.2) node {$q_N$};
					\end{tikzpicture}
\caption{The path integral representation of $\rho_{{N_0^{bath}}}$ and   $\rho_{{N_0^{bath}}'}$ which are the semiclassical limit of $\bm{\rho}_{{N_0^{bath}}}$ and   $\bm{\rho}_{{ N_0^{bath}}'}$ respectively. \label{fig:Dpathintegral2}} 
\end{figure}


 The operator $ \mathbb{U}_{bath}(t,u)$ can be identified with the operator $\mathbb{B}(u)$ provided, in the semiclassical limit, it has the action that matches with the second assertion of the island paradigm. In the remaining sections of the paper  we will show via the gravitational path integral analysis that the action of $ \mathbb{U}_{bath}(t,u)$ on a local operator in the bath matches with the second assertion of the island paradigm. \par


\section{Gravitational path integral and the action of  \texorpdfstring{$\mathbb{U}_{bath}(t,u)$}{Lg} }\label{sect 3}

In this section,  we will investigate the action of  $\mathbb{U}_{bath}(t,u)$ on a local  operator $\mathcal{O}$ in the semiclassical limit by studying the $G_N\to 0$ limit of the correlation functions of the type 

\begin{equation}
\label{pathrep}
A\left(t,u\right)= \langle \bm{\Omega}| \mathbb{U}_{bath}^{\dagger}(t,u)\mathcal{O} \mathbb{U}_{bath}(t,u)  |\bm{\Omega}\rangle.
\end{equation}
  where $|\bm{\Omega}\rangle$  is a quantum JT gravity state whose semiclassical limit $ |\Omega\rangle$ is the Hartle-Hawking state obtained by performing the CFT path integral over the lower half of the Euclidean $w$-plane.  The correlation function $A(t,u)$  can be obtained from the following correlation function
\begin{equation}\label{necorrel}
A\left(n_1, n_2,u\right)=
 \langle \bm{\Omega}| \bm{\rho}^{n_1}_{{N_{u}^{bath}}'} \bm{\rho}_{N^{bath}_{u-a}}^{n_2}\bm{\rho}^{n_2}_{{M_u^{bath}}'}\bm{\rho}_{{M_u^{bath}}}^{n_1}\mathcal{O} \bm{\rho}^{n_1}_{{M_u^{bath}}'} \bm{\rho}_{{M_u^{bath}}}^{n_2}\rho^{n_2}_{{N_{u}^{bath}}'} \bm{\rho}_{{N_{u-a}^{bath}}}^{n_1} |\bm{\Omega}\rangle
\end{equation}
 after a Euclidean continuation $n_1 \to it, ~n_2 \to -it$, this becomes the correlation function of our interest. For integer values of  $n_1$ and $n_2$, in the $G_N\to 0$ limit before the Page time, this correlation function is made using $4n_1+4n_2$ density operators, each of which can be described using the CFT path integral over a $w$-plane with a suitable cut, see figures \ref{fig:Dpathintegral} and \ref{fig:Dpathintegral2}. We will denote this semiclassical limit of  the above density matrices as ${\rho}_{{M_u^{bath}}},{\rho}_{{M_u^{bath}}'}, {\rho}_{{N_{u-a}^{bath}}}$ and ${\rho}_{{N_{u}^{bath}}'}$. Therefore, in this limit the \eqref{necorrel} correlation function can be understood as a  path integral of the CFT defined on surface $\mathcal{S}^T_{4n_1+4n_2+1}$ multiplied by the exponential of the JT gravity action 
\begin{equation}\label{necorrelTP}
A\left(n_1,n_2,u\right)=e^{-S_{JT}\left(\mathcal{S}^T_{4n_1+4n_2+1}\right)}\langle \mathcal{O}  \rangle_{\mathcal{S}^T_{4n_1+4n_2+1}}
 \end{equation}
where $\mathcal{S}^T_{4n_1+4n_2+1}$ is a Riemann surface 
constructed by sewing the $4n_1+4n_2$ copies of the $w$-planes representing the density matrices to the base $w$-plane, where the operator $\mathcal{O}$ is inserted, as shown in figure \ref{fig:USinpathintegral}.
  \begin{figure}
\centering
		\begin{tikzpicture}[scale=.75]
			\fill[draw=black,fill=cyan!5] (-2.25,-2.25) rectangle (2.25,2.25);
			\fill[draw=black, fill=teal!30] (0,0) circle[radius=1];
			\draw[blue, very thick] (1,0.02) to[out=10, in=150] (2.25,0);
			\draw[cyan,very thick] (-1,-0.02) to[out=-160, in=-30] (-2.25,0);
			\draw[red, very thick] (1,-0.02) to[out=-10, in=-150] (2.25,0);
			\draw[orange, very thick] (-1,0.02) to[out=160, in=30] (-2.25,0);
			\draw[brown, very thick] (1.55,0) to[out=30, in=120] (2.25,0);
			\draw[yellow, very thick] (-1,0) to[out=-150, in=-60] (-2.25,0);
			\draw[green, very thick]  (1.55,0) to[out=-30, in=-120] (2.25,0);
			\draw[yellow, very thick] (-1,0) to[out=150, in=60] (-2.25,0);
			\draw (1.95,0) node [circle,fill,inner sep=1pt]{};
			\draw (2,0) node [below]{\texorpdfstring{$\mathcal{O}(p)$}{Lg} };
			\fill[draw=black,fill=cyan!5] (-3,2) rectangle (-5,0);
			\fill[draw=black, fill=teal!30] (-4,1) circle[radius=0.5];
			\draw[blue, very thick] (-3,1.02) -- (-3.5,1);
			\draw (-2.5,1.2) node {\texorpdfstring{   $\rho_{{M_u^{bath}}}$}{Lg}};
			\fill[draw=black,fill=cyan!5] (-5.5,2) rectangle (-7.5,0);
			\fill[draw=black, fill=teal!30] (-6.5,1) circle[radius=0.5];
			\draw[cyan,very thick] (-7,1.02) --(-7.5,1);
			\draw (-8,1.2) node {\texorpdfstring{  $\rho_{{M_u^{bath}}'}$}{Lg} };
			\fill[draw=black,fill=cyan!5] (-3,4.5) rectangle (-5,2.5);
			\fill[draw=black, fill=teal!30] (-4,3.5) circle[radius=0.5];
			\draw[blue, very thick] (-3,3.52) -- (-3.5,3.5);
			\draw (-2.5,3.7) node {\texorpdfstring{  $\rho_{{M_u^{bath}}}$}{Lg} };
			\fill[draw=black,fill=cyan!5] (-5.5,4.5) rectangle (-7.5,2.5);
			\draw[color=black, fill=teal!30](-6.5,3.5) circle[radius=0.5];
			\draw[cyan,very thick] (-7,3.52)-- (-7.5,3.5);
			\draw (-8,3.7)  node {\texorpdfstring{  $\rho_{{M_u^{bath}}'}$}{Lg} };
			\fill[draw=black,fill=cyan!5] (5.5,2) rectangle (7.5,0);
			\fill[draw=black, fill=teal!30] (6.5,1) circle[radius=0.5];
			\draw[red, very thick] (7,1.02) --(7.5,1);
			\draw (7.5,1.2) node [right]{\texorpdfstring{  $\rho_{{M_u^{bath}}}$}{Lg} };
			\fill[draw=black,fill=cyan!5] (3,2) rectangle (5,0);
			\fill[draw=black, fill=teal!30] (4,1) circle[radius=0.5];
			\draw[orange, very thick] (3,1.02) --(3.5,1);
			\draw (2.5,1.2) node {\texorpdfstring{ $\rho_{{M_u^{bath}}'}$}{Lg} };
			\fill[draw=black,fill=cyan!5] (5.5,4.5) rectangle (7.5,2.5);
			\draw[color=black, fill=teal!30](6.5,3.5) circle[radius=0.5];
			\draw[red, very thick] (7,3.52) --  (7.5,3.5);
			\draw (7.5,3.7) node [right]{\texorpdfstring{ $\rho_{{M_u^{bath}}}$}{Lg} };
			\fill[draw=black,fill=cyan!5] (3,4.5) rectangle (5,2.5);
			\fill[draw=black, fill=teal!30] (4,3.5) circle[radius=0.5];
			\draw[orange, very thick] (3,3.52) --(3.5,3.5);
			\draw (2.5,3.7) node {\texorpdfstring{$\rho_{{M_u^{bath}}'}$}{Lg} };
			\fill[draw=black,fill=cyan!5] (-3,7.5) rectangle (-5,5.5);
			\fill[draw=black, fill=teal!30] (-4,6.5) circle[radius=0.5];
			\draw[brown, very thick] (-3.4,6.5)-- (-3,6.5);
			\draw (-3,6.7) node [right]{\texorpdfstring{$\rho_{{N_{u-a}^{bath}}}$}{Lg} };
			\fill[draw=black,fill=cyan!5] (-5.5,7.5) rectangle (-7.5,5.5);
			\draw[color=black, fill=teal!30](-6.5,6.5) circle[radius=0.5];
			\draw[yellow, very thick] (-7.5,6.5) -- (-7,6.5);
			\draw (-8,6.7) node {\texorpdfstring{$\rho_{{N_{u-a}^{bath}}'}$}{Lg}  };
			\fill[draw=black,fill=cyan!5] (-3,10) rectangle (-5,8);
			\fill[draw=black, fill=teal!30] (-4,9) circle[radius=0.5];
			\draw[brown, very thick] (-3.4,9) -- (-3,9);
			\draw (-3,9.2) node [right]{\texorpdfstring{$\rho_{{N_u^{bath}}}$}{Lg}   };
			\fill[draw=black,fill=cyan!5] (-5.5,10) rectangle (-7.5,8);
			\draw[color=black, fill=teal!30](-6.5,9) circle[radius=0.5];
			\draw[yellow, very thick] (-7.5,9) -- (-7,9);
			\draw (-8,8.9) node {\texorpdfstring{$\rho_{{N_u^{bath}}'}$}{Lg}  };
			\fill[draw=black,fill=cyan!5] (5.5,7.5) rectangle (7.5,5.5);
			\draw[color=black, fill=teal!30](6.5,6.5) circle[radius=0.5];
			\draw[green, very thick] (7.1,6.5) -- (7.5,6.5);
			\draw (7.5,6.7) node [right]{\texorpdfstring{$\rho_{{N_u^{bath}}}$}{Lg}  };
			\fill[draw=black,fill=cyan!5] (3,7.5) rectangle (5,5.5);
			\fill[draw=black, fill=teal!30] (4,6.5) circle[radius=0.5];
			\draw[yellow, very thick] (3.5,6.5)-- (3,6.5);
			\draw (2.5,6.7)  node { \texorpdfstring{$\rho_{{N_u^{bath}}'}$}{Lg} };
			\fill[draw=black,fill=cyan!5] (5.5,10) rectangle (7.5,8);
			\draw[color=black, fill=teal!30](6.5,9) circle[radius=0.5];
			\draw[green, very thick] (7.1,9) --(7.5,9);
			\draw (7.5,9.2) node [right]{ \texorpdfstring{$\rho_{{N_u^{bath}}}$}{Lg} };
			\fill[draw=black,fill=cyan!5] (3,10) rectangle (5,8);
			\fill[draw=black, fill=teal!30] (4,9) circle[radius=0.5];
			\draw[yellow, very thick] (3.5,9) --(3,9);
			\draw (2.5,8.9)  node { \texorpdfstring{$\rho_{{N_u^{bath}}'}$}{Lg} };
		\end{tikzpicture}
\caption{The trivial saddle is obtained by cyclically gluing the  \texorpdfstring{$w$}{Lg}-planes along the cuts having same colours.  It is a sheeted geometry over the $w$-plane  having  branch points  at locations \texorpdfstring{$b$}{Lg}, \texorpdfstring{$q_M$}{Lg},\texorpdfstring{$q_N$}{Lg}  on the \texorpdfstring{$w$}{Lg}-plane. The figure describes the gluing for \texorpdfstring{$n_1=n_2=2$}{Lg}.  \label{fig:USinpathintegral}}
\end{figure}
For finite values of $G_N$, the path integral representation of the correlation function \eqref{pathrep} can be expressed as follows  
\begin{align} \label{necorrelp}
&\langle  \bm{\Omega}|{\mathbb{U}_{bath}(s,u)}^{\dagger} \mathcal{O}~\mathbb{U}_{bath}(s,u) |\bm{\Omega} \rangle\nonumber\\
&=  \sum_{\substack{g_i, m_i\\  \sum_ig_i=g\quad\\ \sum_i n_i=4n_1+4n_2+1}}\int_{\otimes_i \mathcal{M}_{g_i,m_i}}\prod_i \prod_{j=1}^{3g_i+3-m_i} dl^i_j d\theta^i_j ~e^{-S_{JT}\left(\{l^i_j,\theta^i_j\}\right)} \langle \mathcal{O} \rangle_{\mathcal{S}_{g,4n_1+4n_2+1}}|_{\substack{n_1 = it\\n_2 = -it}}
\end{align}
where $\mathcal{M}_{g_i,m_i}$ is the moduli space of genus $g_i$ hyperbolic Riemann surface $\mathcal{R}_{g_i, m_i}$ with $m_i$ boundaries.   A point in $\otimes_i\mathcal{M}_{g_i,m_i}$ corresponds  to a  collection of surfaces $\left\{\mathcal{R}_{g_i, m_i}\right\}$. It is  parameterised using  the Fenchel-Nielsen coordinates  $\left(l^i_1,\theta_{1}\cdots, l^i_{3g_i-3+m_i},\theta_{3g_i-3+m_i}\right)$.  The JT gravity action  $S_{JT}\left(\{l^i_j,\theta^i_j\}\right)$ is a function of these moduli parameters . The CFT correlation function $\langle  \mathcal{O} \rangle_{\mathcal{S}_{g,4n_1+4n_2+1}}$ is evaluated on the two dimensional surface $\mathcal{S}_{g,4n_1+4n_2+1}$ constructed by attaching these collection of Riemann surfaces to the boundaries of another surface $\mathcal{B}_{ 4n_1+4n_2+1}$. The surface $\mathcal{B}_{ 4n_1+4n_2+1}$ is obtained by removing the gravitational region from $\mathcal{S}^T_{4n_1+4n_2+1}$.

In this notation, $\mathcal{S}^T_{4n_1+4n_2+1}$ is an example of $\mathcal{S}_{g, 4n_1+4n_2+1}$ obtained by attaching $4n_1+4n_2+1$  hyperbolic disks to the $4n_1+4n_2+1$ boundaries of $\mathcal{B}_{ 4n_1+4n_2+1}$. We say $\mathcal{S}^T_{4n_1+4n_2+1}$ provides us with a trivial saddle for \eqref{necorrelp}, which is the leading saddle at early Lorentzian times. Below we will see that trivial saddle $\mathcal{S}^T_{ 4n_1+4n_2+1} $ has linearly increasing free energy with respect to the boundary time.   Thus we will ask, if there exists another saddle with lower energy, so that we can approximate \eqref{necorrelp} with this new saddle. We will construct explicitly a wormhole saddle $\mathcal{S}_{4n_1+4n_2+1}^W$ that has constant free energy by gluing a sphere  $\mathcal{R}_{ 2n_1+2n_2+1}$  with $2n_1+2n_2+1$ boundaries and $2n_1+2n_2$ hyperbolic disks to $\mathcal{B}_{4n_1+4n_2+1}$ along their boundaries.  The precise geometry of the sphere $\mathcal{R}_{ 2n_1+2n_2+1}$ can be identified by minimising the  free energy of the wormhole geometries $\mathcal{S}_{4n_1+4n_2+1}$. 

\section{Construction of wormhole saddle} \label{sect 4}

In this section, we will compute the free energy of the trivial gravitational saddle that contributes to  \eqref{necorrelp} and construct a wormhole saddle having less free energy after the Page time.  The computation of the free energy of a gravitational saddle of JT gravity coupled to a CFT path integral involves the evaluation of the JT gravity action and the partition function of the CFT defined on the saddle geometry.  Since the phenomena that we are interested to study is not expected to be sensitive to the  fine details of the matter CFT, we shall only  compute  the universal terms in the free energy. \par

 The free energy of a CFT defined on a space having characteristic length $L$ has the behaviour \cite{Cardy:1988tk}
 \begin{equation}\label{F}
     F = A L^2 + BL + C +c (-\frac{\chi}{6} +\Theta(\theta)+ \Phi(\phi))\ln L +  O(1).
 \end{equation}
The terms $AL^2, BL$ are due to bulk and boundary contributions respectively, with $A, B$ being theory dependent constants, $\chi$ is the Euler characteristic and  $C$ is a universal constant. The functions  $\Theta(\theta)$ and $ \Phi(\phi)$ are also theory independent. The function $\Theta(\theta)$  is associated to a conical singularity and is given by
 \begin{equation} \label{conical}
     \Theta(\theta)= \frac{(\theta+2\pi)}{24\pi}\left( 1-\left(\frac{2\pi}{\theta+2\pi} \right)^2\right),
 \end{equation}
where $\theta$ is the conical excess angle. The function $\Phi(\phi)$  is associated to a corner and has the following form
  \begin{equation} \label{corner}
\Phi(\phi)= \frac{\phi}{24\pi}\left( 1-\left(\frac{\pi}{\phi} \right)^2\right)
 \end{equation}
where $\phi$ is the interior angle of the corner.   In this subsection, we will apply this result to discuss the time dependence of the free energy of the trivial saddle,  which is a conformally flat geometry with conical singularities.

\subsection{Trivial saddle}

Following \cite{Cardy:1988tk}, the characteristic length for trivial saddle that gives rise to a time dependent free energy after Lorentzian continuation is given by the Euclidean distance between the  conical singularities at points $b_M^R$ and $b_M^L$: 
  \begin{equation}
\label{charlength}
L=\sqrt{(w_{q_M}-w_{b})(\bar w_{q_M}-\bar w_{b})} .
\end{equation}
 Further, the geometry describing the trivial saddle has eight conical singularities and we need to sum over the contribution from all of them. Four of them have excess angle $\gamma=2\pi n_1$ and the other four have $\gamma=2\pi n_2$. That makes the total contribution from conical singularities to be
\begin{equation}
 2 c [(\Theta(2\pi n_1) + \Theta(2\pi n_2 )]\cdot \log \left((w_{q_M}-w_{b})(\bar w_{q_M}-\bar w_{b})\right)
\end{equation}
This is the universal time dependent part of the free energy for the trivial saddle $F_{T}$, independent of the details of the CFT, except for the central charge.
Doing a Lorentzian continuation to a large time $u$, it becomes
\begin{equation}\label{trivialsaddleFE}
 F_{T}\approx \frac{8\pi c }{\beta} [(\Theta(2\pi n_1) + \Theta(2\pi n_2 )]\cdot u.
\end{equation}

This unbounded growth makes it clear that at late times some other saddle with constant free energy will dominate the trivial saddle, and below we will show there is such a saddle.


\subsection{Wormhole geometry}

 A class of geometries that are summed over in the gravitational path integral \eqref{necorrelp} can be obtained by gluing an arbitrary hyperbolic sphere $\mathcal{R}_{2n_1+2n_2+1}$ and $2n_1+2n_2$   hyperbolic disks to $\mathcal{B}_{4n_1+4n_2+1}$.  We will  denote such a wormhole as $\mathcal{S}^W_{ 4n_1+4n_2+1}$ and detail its construction below.

  \begin{figure}
\centering
 	\begin{tikzpicture}[scale=.7]
          \draw[teal!30,fill=teal!20]plot [smooth] coordinates {(-3.25,12)(-2,9)(0,8)(2,9)(3.25,12)(3.4,12.5)(4,12.75)(4.5,12.55)(4.75,12.2)(4.7,11.65)(4.7,11.65)(6.15,10.75)(4.5,9)     
           (2,5.5)(1.5,.25) (1.2,.5)(1,0)(-1,0)(-1.5,.25)(-2,5.5)(-4.5,9)(-6.15,10.75) (-4.75,12.2)(-3.4,12.5)(-3.25,12)} ;           
			\fill[draw=black, fill=white!20] (0,0) circle[radius=1.5];
			\fill[draw=black, fill=white!20] (5,11.75) circle[radius=1.5];
			\fill[draw=black, fill=white!20] (-5,11.75) circle[radius=1.5];
            \draw [black]plot [smooth] coordinates {(1.5,.25)(2,5.5)(4.5,9)(6.15,10.75)};
              \draw [black]plot [smooth] coordinates {(-1.5,.25)(-2,5.5)(-4.5,9)(-6.15,10.75)};
                 \draw [black]plot [smooth] coordinates {(3.5,12)(2,9)(0,8)(-2,9)(-3.5,12)} ;
                  \draw [black,thick]plot [smooth] coordinates {(0.5,6)(.4,7)(0,8)} ;
                   \draw [black,thick, dashed]plot [smooth] coordinates {(0,8)(-.4,7)(-.5,6)(-.5,5.7)} ;
                    \draw [black,thick]plot [smooth] coordinates {(0.5,6)(1.5,5.6)(2,5)} ;
                   \draw [black,thick, dashed]plot [smooth] coordinates {(-.5,5.7)(1,5.3)(2,5)} ;
                      \draw [black,thick]plot [smooth] coordinates {(0.5,6)(-1,5.1)(-2,5)} ;
                   \draw [black,thick, dashed]plot [smooth] coordinates {(-.5,5.7)(-1,5.3)(-2,5)} ;         
                \end{tikzpicture}
\caption{A hyperbolic pair of pants made by gluing 6 hyperbolic quadrilaterals. As we consider gravitational path integrals, we will require that there be no conical singularities at the contact point, which means that sum of angles should sum to $2\pi$, which will fix the geometry. \label{fig:USinpathintegralwormholeswmapa}} 
\end{figure}


   \begin{figure}
\centering
 	\begin{tikzpicture}[scale=.7]
	 \draw[teal!30,fill=teal!20]plot [smooth] coordinates {(-10,6.1) (-10,1)(-10,0)(-10.5,0)(-12.5,0)(-13,0)(-13,0.5) (-13,6.1)(-12.5,5)(-11.5,4.5)(-10.5,5.1)(-10,6.1) } ; 
	  \draw[teal!30,fill=teal!20]plot [smooth] coordinates {(-5,6.1) (-5,1)(-5,0)(-5.5,0)(-7.5,0)(-8,0)(-8,0.5) (-8,6.1)(-7.5,5)(-6.5,4.5)(-5.5,5.1)(-5,6.1) } ; 
	  \draw[teal!30,fill=teal!20]plot [smooth] coordinates {(-1.5,0)(-.75,3.8) (-.75,3.8) (-.5,4)(-.35,4.5) (-.3,6.1)(.15,4.5)(.5,4)(.75,3.8)(.75,3.8)(1.5,0)(0,1.5) (-1.5,0)} ; 
	   \draw[teal!30,fill=teal!20]plot [smooth] coordinates {(-.75,3.8)(-.75,4)(-.5,4.5)(0.3,6.1)(.6,4.5)(.75,4)(.75,3.8)(-.75,3.8)} ;
	\draw[  thick,middlearrow={>}]plot [smooth] coordinates{(-10,6.1) (-10,0)};
	\draw[ middlearrow={<}, thick]plot [smooth] coordinates{(-13,6.1) (-13,0)};
	\draw[ thick]plot [smooth] coordinates{(-13,0) (-10,0)};
	\draw[  thick,middlearrow={<}]plot [smooth] coordinates{(-8,6.1) (-8,0)};
	\draw[ thick]plot [smooth] coordinates{(-13,6.1) (-12.5,5)(-11.5,4.5)(-10.5,5.1)(-10,6.1)};
	\draw[ thick]plot [smooth] coordinates{(-8,6.1) (-7.5,5)(-6.5,4.5)(-5.5,5.1)(-5,6.1)};
	\draw[  thick]plot [smooth] coordinates{(-8,0) (-5,0)};
	\draw[ middlearrow={>}, thick]plot [smooth] coordinates{(-5,6.1) (-5,0)};
	\draw[  thick,->]plot [smooth] coordinates{(-4.5,3) (-2,3)};
			\fill[draw=black, thick, fill=white!20] (0,0) circle[radius=1.5];
			\draw (-0.3,6.1)  node  [circle,fill,inner sep=1pt]{};
			\draw (0.3,6.1)  node  [circle,fill,inner sep=1pt]{};
			\draw (1.5,0) node  [circle,fill,inner sep=1pt]{};
			\draw (1.5,0.2) node [right]{$P$};
			\draw (-1.5,0) node  [circle,fill,inner sep=1pt]{};
			\draw (-2.6,0.2) node [right]{$Q$};
			\draw (-0.3,6.1) node [left]{$R$};
			\draw (0.3,6.1) node [right]{$S$};
			\draw (-5,0.2) node [right]{$P'$};
			\draw (-5,6.2) node [right]{$R'$};
			\draw (-8,0.2) node [left]{$Q'$};
			\draw (-8,6.2) node [left]{$S'$};
			\draw (-10,0.2) node [right]{$P''$};
			\draw (-10,6.2) node [right]{$R''$};
			\draw (-13,0.2) node [left]{$Q''$};
			\draw (-13,6.2) node [left]{$S''$};
			\draw[->]plot [smooth] coordinates{(-5.3,5.5) (-5.15,5.4) (-5,5.4)};
			\draw (-5.6,5.1) node [right]{$\alpha$};
			\draw[<-]plot [smooth] coordinates{(-8,5.2) (-7.8,5.3) (-7.7,5.4)};
			\draw (-8.2,4.9) node [right]{$\alpha$};
			\draw[->]plot [smooth] coordinates{(-10.3,5.5) (-10.15,5.4) (-10,5.4)};
			\draw (-10.6,5.1) node [right]{$\alpha$};
			\draw[<-]plot [smooth] coordinates{(-13,5.2) (-12.8,5.3) (-12.7,5.4)};
			\draw (-13.2,4.9) node [right]{$\alpha$};
         \draw [black, densely dotted, very thick]plot [smooth] coordinates {(-.3,6.1)(.15,4.5)(.5,4)(.75,3.8)} ;
         \draw [black, thick]plot [smooth] coordinates {(0.3,6.1)(.6,4.5)(.75,4)(.75,3.8)} ;
         \draw [black, densely dotted, very thick]plot [smooth] coordinates {(-.3,6.1)(-.35,4.5)(-.5,4)(-.75,3.8)} ;
         \draw [black, thick]plot [smooth] coordinates {(0.3,6.1)(-.5,4.5)(-.75,4)(-.75,3.8)} ;
                  \draw [black,thick]plot [smooth] coordinates {(.75,3.8)(1.5,0)} ;
         \draw [black,thick]plot [smooth] coordinates {(-1.5,0)(-.75,3.8)} ;
                         \end{tikzpicture}
\caption{ The construction of a symmetric crown using two identical hyperbolic quadrilaterals. The name crown is due to the shape of the resulting geometry, in which quadrilaterals have a flat base and pointed vertices at the top. Here $2\alpha$ is the ``crown angle", which uniquely identifies the symmetric crown's shape. \label{fig:USinpathintegralwormholeswmap12}} 
\end{figure}

   \begin{figure}
\centering
 	\begin{tikzpicture}[scale=1]
	  \draw[teal!30,fill=teal!20]plot [smooth] coordinates {(-1.5,0)(-.75,3.8) (-.75,3.8) (-.5,4)(-.35,4.5) (-.3,6.1)(.15,4.5)(.5,4)(.75,3.8)(.75,3.8)(1.5,0)(0,1.5) (-1.5,0)} ; 
	   \draw[teal!30,fill=teal!20]plot [smooth] coordinates {(-.75,3.8)(-.75,4)(-.5,4.5)(0.3,6.1)(.6,4.5)(.75,4)(.75,3.8)(-.75,3.8)} ;
	       \fill[draw=black,fill=cyan!5] (-2.5,2.5) rectangle (+2.5,-2.5);
				\fill[draw=black, thick, fill=white!20] (0,0) circle[radius=1.5];
			\draw[->, thick]plot [smooth] coordinates{(.45,5.4) (.35,5.2) (.15,5.2)  (0,5.4)};
			\draw (-.2,6.5) node [right]{$2\alpha$};
			\draw (-1,5) node [right]{$\tilde e$};
			\draw (.5,5) node [right]{$ e$};
         \draw [black, densely dotted, very thick]plot [smooth] coordinates {(-.3,6.1)(.15,4.5)(.5,4)(.75,3.8)} ;
         \draw [black, thick]plot [smooth] coordinates {(0.3,6.1)(.6,4.5)(.75,4)(.75,3.8)} ;
         \draw [black, densely dotted, very thick]plot [smooth] coordinates {(-.3,6.1)(-.35,4.5)(-.5,4)(-.75,3.8)} ;
         \draw [black, thick]plot [smooth] coordinates {(0.3,6.1)(-.5,4.5)(-.75,4)(-.75,3.8)} ;
                  \draw [black,thick]plot [smooth] coordinates {(.75,3.8)(1.5,0)} ;
         \draw [black,thick]plot [smooth] coordinates {(-1.5,0)(-.75,3.8)} ;
                         \end{tikzpicture}
\caption{ The crown attached to the bath can be identified as a $w$-plane with a cut. The deficit angle at the end points of the cut is given by $2\alpha-2\pi$. The two edges of the crown are denoted as $e$ and $\tilde e$.  \label{fig:USinpathintegralwormholeswmap121}}
\end{figure}

\begin{figure}
\centering
 	\begin{tikzpicture}[scale=1]
   	\fill[draw=black,fill=cyan!5] (-2.5,-2.5) rectangle (2.5,2.5);
			\fill[draw=black, fill=white!20] (0,0) circle[radius=1.5];
			\draw[blue, very thick,middlearrow={>}] (1.5,0.02) to[out=10, in=150] (2.5,0);
			\draw[cyan,very thick,middlearrow={<}] (-1.5,-0.02) to[out=-160, in=-30] (-2.5,0);
			\draw[olive, very thick,middlearrow={<}] (1.5,-0.02) to[out=-10, in=-150] (2.5,0);
			\draw[orange, very thick,middlearrow={>}] (-1.5,0.02) to[out=160, in=30] (-2.5,0);
			\draw[brown, very thick,middlearrow={<}] (1.8,0) to[out=30, in=120] (2.5,0);
			\draw[yellow, very thick,middlearrow={>}] (-1.5,0) to[out=-150, in=-60] (-2.5,0);
			\draw[green, very thick,middlearrow={<}]  (1.8,0) to[out=-30, in=-120] (2.5,0);
			\draw[violet, very thick,middlearrow={>}] (-1.5,0) to[out=150, in=60] (-2.5,0);
			\draw (1.5,0) node  [circle,fill,inner sep=1pt]{};
			\draw (1.95,0.1) node [above]{$q_{{N}}$};
			\draw (1.75,0) node  [circle,fill,inner sep=.75pt]{};
			\draw (-1.5,0) node  [circle,fill,inner sep=.7pt]{};
			\draw (2.2,0) node [circle,fill,inner sep=1pt]{};
			\draw (2.25,0) node [below]{$\mathcal{O}(p)$};
			\draw (.5,-0.2) node [right]{$q_M$};
			\draw (-1.5,0) node  [circle,fill,inner sep=1pt]{};
			\draw (-.9,-0.2) node [left]{$b$};
         \draw [black,dashed,thick]plot [smooth] coordinates {(-2.5,7.6)(-1.7,7)(-1,7)(-.8,8.2)} ;
         \draw [black,thick]plot [smooth] coordinates {(-2.5,7.3)(-1.7,6.7)(-.9,6.7)(-.8,8.2)} ;
           \draw [black,dashed,thick ]plot [smooth] coordinates {(2.5,7.6)(2.7,7)(3.9,6.4)(5.5,6)} ;
         \draw [black,thick]plot [smooth] coordinates {(2.5,7.3)(2.7,6.7)(3.9,6.3)(5.5,6)} ;
          \draw [black,dashed,thick ]plot [smooth] coordinates {(-2.5,7.6)(-2.7,7)(-3.9,6.4)(-5.5,6)} ;
         \draw [black,thick]plot [smooth] coordinates {(-2.5,7.3)(-2.7,6.7)(-3.9,6.3)(-5.5,6)} ;
         \draw [black,dashed,thick ]plot [smooth] coordinates {(-2.5,7.6)(-1.7,7)(-1,7)(-.8,8.2)} ;
         \draw [black,thick]plot [smooth] coordinates {(-2.5,7.3)(-1.7,6.7)(-.9,6.7)(-.8,8.2)} ;
           \draw [black,dashed,thick ]plot [smooth] coordinates {(2.5,7.6)(1.7,7)(1,7)(.8,8.2)} ;
         \draw [black,thick]plot [smooth] coordinates {(2.5,7.3)(1.7,6.7)(.9,6.7)(.8,8.2)} ;
         \draw [black,dashed,thick]plot [smooth] coordinates {(2.5,5.2)(.9,5.4)(.5,4.2)(.6,3.5)(.85,3.4)} ;
         \draw [black,thick]plot [smooth] coordinates {(2.3,4.8)(1.1,5.1)(.65,4)(.85,3.4)} ;
         \draw [black,dashed,thick]plot [smooth] coordinates {(-2.3,5.2)(-.9,5.4)(-.5,4.2)(-.6,3.5)(-.85,3.4)} ;
         \draw [black,thick]plot [smooth] coordinates {(-2.3,4.8)(-1.1,5.1)(-.65,4)(-.85,3.4)} ;
           \draw [black,dashed,thick]plot [smooth] coordinates {(2.5,5.2)(2.9,5.7)(3.6,5.9)(4.4,5.8)(5,5.5)} ;
         \draw [black,thick]plot [smooth] coordinates {(2.3,4.8)(3.1,5.6)(4.2,5.8)(5,5.5)} ;
                  \draw [black,dashed,  thick]plot [smooth] coordinates {(-2.5,5.2)(-2.9,5.7)(-3.6,5.9)(-4.4,5.8)(-5,5.5)} ;
         \draw [black,thick]plot [smooth] coordinates {(-2.3,4.8)(-3.1,5.6)(-4.2,5.8)(-5,5.5)} ;
         \draw (-2.3,4.8)  node  [circle,fill,inner sep=1pt]{};
          \draw (2.3,4.8)  node  [circle,fill,inner sep=1pt]{};
           \draw (-2.5,5.2)  node  [circle,fill,inner sep=1pt]{};
          \draw (2.5,5.2)  node  [circle,fill,inner sep=1pt]{};
              \draw (-2.5,7.3)  node  [circle,fill,inner sep=1pt]{};
          \draw (2.5,7.3)  node  [circle,fill,inner sep=1pt]{};
           \draw (-2.5,7.6)  node  [circle,fill,inner sep=1pt]{};
          \draw (2.5,7.6)  node  [circle,fill,inner sep=1pt]{};      
            \draw (-2.9,4.9)  node {$j_{{N}'}$};
          \draw (-2.2,4.3)  node {$j_{{N}}$};
             \draw (3,4.9)  node {$\tilde j_{{N}'}$};
          \draw (2.3,4.3)  node {$\tilde j_{{N}}$};        
           \draw (2.9,7.9)  node {$j_{{M}'}$};
          \draw (3,7.2)  node {$j_{{M}}$};
             \draw (-3,8.1)  node {$\tilde j_{{M}'}$};
          \draw (-3.1,7.3)  node {$\tilde j_{{M}}$};
          \draw [black,thick]plot [smooth] coordinates {(5.5,6)(5,5.5)} ;
           \draw [black,thick]plot [smooth] coordinates {(-5.5,6)(-5,5.5)} ;
         \draw [black,thick]plot [smooth] coordinates {(.9,3.4)(.7,3)(1.5,0)} ;
         \draw [black,thick]plot [smooth] coordinates {(-1.5,0)(-.7,3)(-.9,3.4)} ;
                 \draw [black,thick]plot [smooth] coordinates {(.9,8.2)(0,8)(-.9,8.2)} ;
       \draw[->, thick]plot [smooth] coordinates{(2.55,6.9) (2.45,6.7) (2.25,6.7)  (2.1,6.9)};
			\draw (1.8,6.25) node [right]{$2\beta$};
                \end{tikzpicture}
\caption{ The tree attached to the bath  is a $w$-plane with four cuts. The four cuts are along the curves   $j_{{N}'}j_{{N}}, \tilde j'_{{N}'} \tilde j_{{N}}, j_{{M}'} j_{{M}}$ and $\tilde j'_{{M}'}\tilde j_{{M}}$. The deficit angle at the end points of half of the cuts is given by $2\beta_1-2\pi$ and the deficit angle at the end points of the remaining half of the cuts is given by $2\beta_2-2\pi$. \label{fig:USinpathintegralwormholeswmap13}} 
\end{figure}

  \begin{figure}
\centering
 	\begin{tikzpicture}[scale=.6]
          \draw[teal!30,fill=teal!20]plot [smooth] coordinates {(-4,9.5) (-2.75,8.35)   (0,8.05)   (3,8.5)(4,9.5)(5.8,12.3)(5.8,12.3)(5.9,12.5)(6.5,12.75)(7,12.55)(7.25,12.2)(7.1,11.55)(7.1,11.55)(6.5,11)(4.85,8.5)   (4.5,6.7)(6.2,7.1)(8.7,8.2)(8.7,8.2)(9,8.25)(9.6,8)(9.75,7.3)(9.5,6.95)(9.1,6.75)(8.9,6.75)(8.9,6.75)(6,6.2)  (5,5.5) (4.55,4.85)  (5,4.5)  (7,3.5)  (9.1,2.75)(9.6,2.5)(9.75,1.8)(9.5,1.45)(9,1.25)(8.7,1.3)(8.7,1.3)(4,4)(2.75,4.25)     (3,2.3)(4.65,0.35)(4.65,0.35)(4.75,-.2)(4.5,-.55)(4,-.75)(3.4,-.5)(3.5,0)(1.5,2.9)(1,3.4)(.7,3)(1.5,0) (1.2,.5)(1,0)(-1,0)(-.7,3)(-1,3.4)(-1.5,2.9)(-3.5,0)(-3.5,0)(-3.4,-.5)(-4,-.75)(-4.5,-.55)(-4.75,-.2)(-4.65,0.35)(-4.65,0.35)(-3,2.3)  (-2.5,3.85) (-2.75,4.25)(-4,4)(-8.7,1.3)(-9,1.25)(-9.5,1.45)(-9.75,1.8)(-9.6,2.5)(-9.1,2.75)(-9.1,2.75)(-4.5,4.5)(-5,5.5)  (-8.9,6.75)(-8.9,6.75)(-9.1,6.75)(-9.5,6.95)(-9.75,7.3)(-9.6,8)(-9,8.25)(-8.7,8.2)(-8.7,8.2)(-6.2,7.1)(-4.5,6.7)     (-4.85,8.5)(-6.5,11)(-7.1,11.55)(-7.1,11.55)(-7.25,12.2)(-7,12.55)(-6.5,12.75)(-5.9,12.5)(-5.8,12.3)(-5.8,12.3)(-4,9.5) } ;          
			\fill[draw=black,fill=cyan!5] (-2.25,-2.25) rectangle (2.25,2.25);
			\fill[draw=black, fill=white!20] (0,0) circle[radius=1.25];
			\draw[brown, very thick,middlearrow={>}] (1.25,0.02) to[out=10, in=150] (2.25,0);
			\draw[cyan,very thick,middlearrow={<}] (-1.25,-0.02) to[out=-160, in=-30] (-2.25,0);
			\draw[green, very thick,middlearrow={<}] (1.25,-0.02) to[out=-10, in=-150] (2.25,0);
			\draw[orange, very thick,middlearrow={>}] (-1.25,0.02) to[out=160, in=30] (-2.25,0);
			\draw[blue, very thick,middlearrow={<}] (1.55,0) to[out=30, in=120] (2.25,0);
			\draw[yellow, very thick,middlearrow={>}] (-1.25,0) to[out=-150, in=-60] (-2.25,0);
			\draw[olive, very thick,middlearrow={<}]  (1.55,0) to[out=-30, in=-120] (2.25,0);
			\draw[violet, very thick,middlearrow={>}] (-1.25,0) to[out=150, in=60] (-2.25,0);
			\draw (1.25,0) node  [circle,fill,inner sep=.7pt]{};
			\draw (1.5,0) node  [circle,fill,inner sep=.75pt]{};
			\draw (-1.25,0) node  [circle,fill,inner sep=.7pt]{};
			\draw (1.95,0) node [circle,fill,inner sep=1pt]{};
			\draw (2,0) node [below]{$\mathcal{O}(p)$};
			\fill[draw=black,fill=cyan!5] (-3,1) rectangle (-5,-1);
			\fill[draw=black, fill=white!20] (-4,0) circle[radius=0.5];
			\draw[blue,very thick] plot [smooth] coordinates{(-3,0)(-3.4,0)};
			\draw  (-7.5,-1.5)  node [right]{$\rho_{{N_{u}^{bath}}'}$};
			\draw  (5.75,-1.5) node [right]{$\rho_{{N_{u}^{bath}}'}$};
			\draw  (-4.5,13.5) node [right]{$\rho_{{M_u^{bath}}'}$};
			\draw (2.75,13.5)  node [right]{$\rho_{{M_u^{bath}}'}$};
			\fill[draw=black,fill=cyan!5] (-5.5,1) rectangle (-7.5,-1);
			\fill[draw=black, fill=teal!20] (-6.5,0) circle[radius=0.5];
                   \draw[cyan,  thick] plot [smooth] coordinates{(-7,0)(-7.5,0)};
			\draw (-4.5,-1.5)  node [right]{$\rho_{{N_{u-a}^{bath}}}$};
			\draw (3.5,-1.5) node [right]{$\rho_{{N_{u-a}^{bath}}}$};
			\draw (-7,13.5) node [right]{$\rho_{{M_u^{bath}}}$};
			\draw (6,13.5) node [right]{$\rho_{{M_u^{bath}}}$};
			\fill[draw=black,fill=cyan!5] (-8,3) rectangle (-10,1);
			\fill[draw=black, fill=white!20] (-9,2) circle[radius=0.5];
			\draw[blue,very thick] plot [smooth] coordinates{((-8.4,2)(-8,2)};
			\draw  (-12,4.5) node [right]{$\rho_{{N_{u}^{bath}}'}$};
			\draw  (10.1,4.5) node [right]{$\rho_{{N_{u}^{bath}}'}$};
			\draw  (10.1,10) node [right]{$\rho_{{M_{u}^{bath}}'}$};
			\draw (-12,10) node [right]{$\rho_{{M_{u}^{bath}}'}$};
			\fill[draw=black,fill=cyan!5] (-8,5.5) rectangle (-10,3.5);
			\draw[color=black, fill=teal!20](-9,4.5) circle[radius=0.5];
			\draw[cyan, thick] plot [smooth] coordinates{(-10,4.5)(-9.5,4.5)};
			\draw (-12,2)  node [right]{$\rho_{{N_{u-a}^{bath}}}$};
			\draw (10,2) node [right]{$\rho_{{N_{u-a}^{bath}}}$};
			\draw  (10,7.5) node [right]{$\rho_{{M_u^{bath}}}$};
			\draw (-12,7.5)  node [right]{$\rho_{{M_u^{bath}}}$};
			\fill[draw=black,fill=cyan!5] (5.5,1) rectangle (7.5,-1);
			\fill[draw=black, fill=teal!20] (6.5,0) circle[radius=0.5];
			\fill[draw=black,fill=cyan!5] (3,1) rectangle (5,-1);
			\fill[draw=black, fill=white!20] (4,0) circle[radius=0.5];
			\draw[olive,very thick] plot [smooth] coordinates{(5,0)(4.6,0)};
		\draw[orange,  thick] plot [smooth] coordinates{(6,0)(5.5,0)};
			\fill[draw=black,fill=cyan!5] (8,5.5) rectangle (10,3.5);
			\draw[color=black,fill=teal!20](9,4.5) circle[radius=0.5];
			\fill[draw=black,fill=cyan!5] (8,3) rectangle (10,1);
			\fill[draw=black, fill=white!20] (9,2) circle[radius=0.5];
				\draw[orange,  thick] plot [smooth] coordinates{(8,4.5)(8.5,4.5)};
			\draw[olive,very thick] plot [smooth] coordinates{(9.6,2)(10,2)};
			\fill[draw=black,fill=cyan!5] (-8,11) rectangle (-10,9);
			\fill[draw=black,fill=teal!20] (-9,10) circle[radius=0.5];
			\fill[draw=black,fill=cyan!5] (-8,8.5) rectangle (-10,6.5);
			\draw[color=black, fill=white!20](-9,7.5) circle[radius=0.5];
			\draw[yellow, very thick] plot [smooth] coordinates{(-9.5,10)(-10,10)};
			\draw[brown, very thick] plot [smooth] coordinates{(-8.5,7.5)(-8,7.5)};
			\fill[draw=black,fill=cyan!5] (-3,13) rectangle (-5,11);
			\fill[draw=black, fill=teal!20] (-4,12) circle[radius=0.5];
			\draw[yellow, very thick] plot [smooth] coordinates{(-4.5,12)(-5,12)};
			\fill[draw=black,fill=cyan!5] (-5.5,13) rectangle (-7.5,11);
			\draw[color=black, fill=white!20](-6.5,12) circle[radius=0.5];
			\draw[brown,  thick] plot [smooth] coordinates{(-5.5,12)(-6,12)};
			\fill[draw=black,fill=cyan!5] (8,11) rectangle (10,9);
			\draw[color=black, fill=teal!20](9,10) circle[radius=0.5];
			\draw[violet, very thick]plot [smooth] coordinates{(8,10)(8.5,10)};
			\fill[draw=black,fill=cyan!5] (8,8.5) rectangle (10,6.5);
			\fill[draw=black, fill=white!20] (9,7.5) circle[radius=0.5];
			\draw[green, very thick,] plot [smooth] coordinates{(10,7.5)( (9.5,7.5)};
			\fill[draw=black,fill=cyan!5] (5.5,13) rectangle (7.5,11);
			\draw[color=black, fill=white!20](6.5,12) circle[radius=0.5];
			\draw[green, very thick] plot [smooth] coordinates{(7,12)(7.5,12)};
			\fill[draw=black,fill=cyan!5] (3,13) rectangle (5,11);
			\fill[draw=black, fill=teal!20] (4,12) circle[radius=0.5];
			\draw[violet, very thick] plot [smooth] coordinates{(3.5,12)(3,12)};
           \draw [black,dashed, very thick ]plot [smooth] coordinates {(2.5,7.6)(2.7,7)(3.9,6.4)(5.5,6)} ;
         \draw [black,very thick]plot [smooth] coordinates {(2.5,7.3)(2.7,6.7)(3.9,6.3)(5.5,6)} ;
          \draw [black,dashed, very thick ]plot [smooth] coordinates {(-2.5,7.6)(-2.7,7)(-3.9,6.4)(-5.5,6)} ;
         \draw [black,very thick]plot [smooth] coordinates {(-2.5,7.3)(-2.7,6.7)(-3.9,6.3)(-5.5,6)} ;
         \draw [black,dashed, very thick ]plot [smooth] coordinates {(-2.5,7.6)(-1.7,7)(-1,7)(-.8,8.2)} ;
         \draw [black,very thick]plot [smooth] coordinates {(-2.5,7.3)(-1.7,6.7)(-.9,6.7)(-.8,8.2)} ;
           \draw [black,dashed, very thick ]plot [smooth] coordinates {(2.5,7.6)(1.7,7)(1,7)(.8,8.2)} ;
         \draw [black,very thick]plot [smooth] coordinates {(2.5,7.3)(1.7,6.7)(.9,6.7)(.8,8.2)} ;
            \draw [dashed,thick]plot [smooth] coordinates {(2.5,7.6)(4.5,8.2)(4.8,8)} ;
         \draw [thick]plot [smooth] coordinates {(2.5,7.3)(4.5,7.8)(4.8,8)} ;
            \draw [dashed,thick]plot [smooth] coordinates {(-2.5,7.6)(-4.5,8.2)(-4.8,8)} ;
         \draw [thick]plot [smooth] coordinates {(-2.5,7.3)(-4.5,7.8)(-4.8,8)} ;
         \draw [black,dashed, very thick]plot [smooth] coordinates {(2.5,5.2)(.9,5.4)(.5,4.2)(.6,3.5)(.85,3.4)} ;
       \draw [black,very thick]plot [smooth] coordinates {(2.3,4.8)(1.1,5.1)(.65,4)(.85,3.4)} ;
         \draw [black,dashed, very thick]plot [smooth] coordinates {(-2.3,5.2)(-.9,5.4)(-.5,4.2)(-.6,3.5)(-.85,3.4)} ;
         \draw [black,very thick]plot [smooth] coordinates {(-2.3,4.8)(-1.1,5.1)(-.65,4)(-.85,3.4)} ;
           \draw [black,dashed, very thick]plot [smooth] coordinates {(2.5,5.2)(2.9,5.7)(3.6,5.9)(4.4,5.8)(5,5.5)} ;
         \draw [black,very thick]plot [smooth] coordinates {(2.3,4.8)(3.1,5.6)(4.2,5.8)(5,5.5)} ;
                  \draw [black,dashed, very thick]plot [smooth] coordinates {(-2.5,5.2)(-2.9,5.7)(-3.6,5.9)(-4.4,5.8)(-5,5.5)} ;
         \draw [black,very thick]plot [smooth] coordinates {(-2.3,4.8)(-3.1,5.6)(-4.2,5.8)(-5,5.5)} ;
         \draw (-2.3,4.8)  node  [circle,fill,inner sep=1pt]{};
          \draw (2.3,4.8)  node  [circle,fill,inner sep=1pt]{};
           \draw (-2.5,5.2)  node  [circle,fill,inner sep=1pt]{};
          \draw (2.5,5.2)  node  [circle,fill,inner sep=1pt]{};    
              \draw (-2.5,7.3)  node  [circle,fill,inner sep=1pt]{};
          \draw (2.5,7.3)  node  [circle,fill,inner sep=1pt]{};
           \draw (-2.5,7.6)  node  [circle,fill,inner sep=1pt]{};
          \draw (2.5,7.6)  node  [circle,fill,inner sep=1pt]{};
         	\draw [dashed,thick]plot [smooth] coordinates {(2.5,5.2)(2.95,4.7)(3,4.3)} ;
         \draw [thick]plot [smooth] coordinates {(2.3,4.8)(2.8,4.3)(3,4.3)} ;
      \draw [dashed,thick]plot [smooth] coordinates {(-2.5,5.2)(-2.95,4.7)(-3,4.3)} ;
         \draw [thick]plot [smooth] coordinates {(-2.3,4.8)(-2.8,4.3)(-3,4.3)} ;
         \draw [black]plot [smooth] coordinates {(3.5,0)(1.5,2.9)(1,3.4)(.7,3)(1.25,0)} ;
         \draw [black]plot [smooth] coordinates {(-1.25,0)(-.7,3)(-1,3.4)(-1.5,2.9)(-3.5,0)} ;
          \draw [black]plot [smooth] coordinates {(-4.4,0.35)(-3,2.3)(-2.5,3.85)(-4,4)(-8.9,1.45)} ;
          \draw [black]plot [smooth] coordinates {(-9.1,2.5)(-4.75,4.5)   (-5,5.5)(-6,6.2)(-8.9,7)  };
           \draw [black]plot [smooth] coordinates {(-8.9,8)(-6.2,7.1)(-4.5,6.7)(-4.85,8.5)(-6.35,11)(-6.85,11.6)};
             \draw [black]plot [smooth] coordinates {(-6.05,12.3)(-4,9.5)(-3,8.5) (-2,8.2) (0,8.1) (1.25,8.1)(3,8.5) (4,9.5)(6.05,12.3)};
       \draw [black]plot [smooth] coordinates {(4.4,0.35)(3,2.3)(2.75,4.25)(4,4)(8.9,1.45)};
                    \draw [black]plot [smooth] coordinates {(9.1,2.5)(4.75,4.5)(5,5.5)(6,6.2)(8.9,7)};
           \draw [black]plot [smooth] coordinates {(8.9,8)(6.2,7.1)(4.5,6.7)(4.85,8.5)(6.35,11)(6.85,11.6)};
                \end{tikzpicture}
\caption{ The wormhole geometry ${\mathcal{S}}_{4n_1+4n_2+1}$ obtained by replacing the gravitational region in the trivial saddle with a hyperbolic sphere having $2n_1+2n_2+1$ boundaries and $2n_1+2n_2$ hyperbolic disks. The  thick coloured lines indicate cuts.  All the cuts with the same colour are being cyclicly glued.   In this figure,  $n_1$ and $n_2$ are  chosen to be 2.  \label{fig:USinpathintegralwormholeswmap6}}
\end{figure}

We can glue a set of hyperbolic quadrilaterals having arbitrary angles at the vertices \cite{imayoshi2012introduction,pan2017finite} to get a hyperbolic wormhole geometry. Typically such a geometry can have conical singularities  where the vertices of quadrilaterals meet, however, in  gravity path integrals singular geometries are not allowed. Thus by requiring there be no conical singularities, we can determine the angles at the vertices of the quadrilaterals.

Let us illustrate this by considering the example of $\mathcal{R}_{3}$ geometry shown in figure \ref{fig:USinpathintegralwormholeswmapa}. It is a  hyperbolic sphere with 3 boundaries.
It is made out of three ``crowns", which are surfaces obtained by cutting along the black curves. Topologically each crown is a disk with a cut, and is made by gluing two identical hyperbolic quadrilaterals. This decomposition is illustrated in figure \ref{fig:USinpathintegralwormholeswmap12}.
The angle subtended by each crown at the joining vertex is $2\alpha$, which is determined by the condition of no conical singularities, requiring the sum of all angles be $2\pi$,  which sets  $\alpha=\pi/3$ for each crown. We shall call the angle $\alpha$ as the crown angle. Thus, this geometry can be obtained by gluing six identical hyperbolic quadrilaterals with crown angle $\pi/3$.


 Using the same idea, we can construct the Riemann spheres $\mathcal{R}_{2n_1+2n_2+1}$ using $2n_1+2n_2$ number of  crowns and a ``tree".   The tree shown in figure \ref{fig:USinpathintegralwormholeswmap13} is topologically is a hyperbolic disk with four cuts. It can be constructed by gluing 14 hyperbolic quadrilaterals.  This decomposition of  $\mathcal{R}_{2n_1+2n_2+1}$ can be seen by slitting it along the black curves shown in  \ref{fig:USinpathintegralwormholeswmap6}. The neighbourhood of a vertex of the thick black curve on $\mathcal{S}_{4n_1+4n_2+1}$ is the union of the corners of $n_i$ crowns and a corner of the tree where $i$ can be $1,2$. Let the interior angle of this corner in the crown in $2\alpha_i$ and the interior angle of this corner in the tree is $2\beta_i$.  To proceed, we need to constrain the crown angle $\alpha_i$ and the tree angle $\beta_i$ by demanding the smoothness condition. The condition of the absence of conical singularity demands that
  \begin{align}\label{alphabeta}
  n_i\alpha_i+\beta_i =\pi, \qquad i=1,2.
  \end{align}
This allows us to describe a set of wormhole geometries free of conical singularities.   

\subsection{Wormhole saddle}\label{wsaddle}

The wormhole saddle $\mathcal{S}^W_{ 4n_1+4n_2+1}$ can be identified by extremising the free energy of the wormhole geometry constructed in the previous subsection.   The  free energy $F_W$ of the wormhole geometry  $\mathcal{S}_{ 4n_1+4n_2+1}$ is  the sum of the JT gravity action and the negative of logarithm of the  partition function $Z\left({\mathcal{S}_{4n_1+4n_2+1}}\right)$ of the matter CFT on $\mathcal{S}_{ 4n_1+4n_2+1}$.  The surface $\mathcal{S}_{ 4n_1+4n_2+1}$, as explained above, is obtained by gluing $\mathcal{R}_{ 2n_1+2n_2+1}$, a smooth hyperbolic Riemann surface, and $2n_1+2n_2$ hyperbolic disks $D_1$, $\cdots$, $D_{2n_1+2n_2}$ to  $\mathcal{B}_{4n_1+4n_2+1}$,  a surface having flat metric and conical singularities on it, along the boundaries.  

The partition function of a CFT on a wormhole geometry  $S_3$, a symmetric pair of pants $\mathcal{R}_3$ with copies of the bath attached to its boundaries, can be computed  by calculating the correlation function of the twist operators inserted on a plane at two points $w_R, w_L$ that corresponds to the fixed points on $\mathcal{R}_3$ under the action of the replica symmetry $\mathbb{Z}_3$. Modulo the Weyl factor, it is given by 
\begin{align}\label{z3replica}
F_{\mathcal{R}_3}\approx -\frac{2c}{3}\text{ln} |w_R-w_L|
\end{align}
This answer does not depend on the details of the CFT except the central charge.  Interestingly, such universal terms in the free energy of a CFT  can be obtained in a slightly different way.  Let us  scissor  $S_3$ along the black curves shown in figure \ref{fig:USinpathintegralwormholeswmapa}. The resulting surface is three copies of a plane with two conical singularities having deficit angle $\frac{4\pi}{3}$.    As shown in \cite{Cardy:1988tk}, the contributions to the conformal field theory free energy arising from the points on the surface having conical excess or deficit angle are universal and are given by \eqref{conical}.  These contributions are computed by choosing a neighbourhood that matches with a staircase geometry. The size $L$ appearing in these formulae is the width of the largest staircase geometry that can be identified in the neighbourhood of such points. Here, we can identify $L$ as the distance between the two conical singularities, $L=|w_R-w_L|$, and $\theta=-\frac{4\pi}{3}$. We shall follow this procedure for extracting the universal part of the free energy of a CFT  defined on $\mathcal{S}_{4n_1+4n_2+1}$. 

 Consider scissoring $\mathcal{S}_{4n_1+4n_2+1}$ along the thick black curves shown in figure \ref{fig:USinpathintegralwormholeswmap6}. We shall denote the resulting surface  as $\mathbb{S}_{4n_1+4n_2+1}$.  The gravitational region of $\mathbb{S}_{4n_1+4n_2+1}$ consists of $2n_1+2n_2$ number of crowns and hyperbolic disks and one tree.  Hence, we can express $ Z\left({\mathcal{S}_{4n_1+4n_2+1}}\right)$ as follows
    \begin{align}\label{Zs2}
 &Z\left({\mathcal{S}_{4n_1+4n_2+1}}\right)\nonumber\\
 &=  \sum_{\substack{e_1,\cdots,e_{2n_1+2n_2+1}\\\tilde e_1,\cdots,\tilde e_{2n_1+2n_2+1}}}   Z\left(\mathbb{S}_{4n_1+4n_2+1};\tilde e_1, e_{1},\cdots,\tilde e_{2n_1+2n_2+4}, e_{2n_1+4n_2+4}\right) \prod_{l=1}^{2n_1+2n_2+4}\langle \tilde e_l|e_l\rangle
 \end{align}
 where, the $e'$s and $\tilde e'$s are the boundary conditions at the edges of the  crowns or tree.

 The connected surface  $\mathbb{S}_{4n_1+4n_2+1}$ is a geometry having $2n_1+2n_2+4$ cuts with end points having conical deficit angle and conical singularities. It can be  described by specifying the geometry of the $4n_1+4n_2+1$ number of sheets that are glued along the cuts in the bath region. These sheets can be grouped as follows:
 \begin{itemize}
 \item    $n_1$ number of $w$-planes with a cut in the gravitational region connecting the points  $i_M$ and  $i_{M'}$  and a cut in the bath region along the interval $\left(q_M,i_0^R\right)$. The deficit angle at the points  $i_M$ and  $i_{M'}$ are $2\pi-2\alpha_1$.
  \item $n_2$ number of  $w$-planes with a cut in the gravitational region connecting the points  $i_N$ and  $i_{N'}$  and a cut in the bath region along the interval $\left(q_N,q_M\right)\cup \left(q_N,i_0^R\right)$. The deficit angle at the points  $i_N$ and  $i_{N'}$ are $2\pi-2\alpha_2$.  
 \item   $n_2$ number of $w$-planes with a cut in the gravitational region connecting the points  $\tilde{i_M}$ and  $\tilde{i_{M'}}$  and a cut in the bath region along the interval $\left(q_M,i_0^R\right)$. The deficit angle at the points  $\tilde{i_M}$ and  $\tilde{i_{M'}}$ are $2\pi-2\alpha_2$.  
 \item  $n_1$ number of $w$-planes with a cut in the gravitational region connecting the points  $\tilde{i_N}$ and  $\tilde{i_{N'}}$  and a cut in the bath region along the interval $\left(q_N,q_M\right)\cup \left(q_N,i_0^R\right)$. The deficit angle at the points  $\tilde{i_N}$ and  $\tilde{i_{N'}}$ are $2\pi-2\alpha_1$. 
 \item    $2n_1+2n_2$ number sheets are  $w$-planes with a cut in the bath region along the interval $\left(b,i_0^L\right)$.
 \item   $w$-plane with four cuts along the intervals $\left(j_{{N}}, j_{{N}'}\right)$, $\left(\tilde j_{{N}}, \tilde j_{{N}'}\right)$, $\left(j_{{M}}, j_{{M}'}\right),$ and $\left(\tilde j_{{M}},\tilde j_{{M}'}\right)$ respectively  in the gravitational region and all the ten cuts described above in the bath region. The deficit angle at the points  $j_{{M}}$, $j_{{M}'}$, $\tilde j_{{N}}$, $\tilde j_{{N}'}$  are $2\pi-2\beta_1$  and $j_{{N}}$, $j_{{N}'}$, $\tilde j_{{M}}$, $\tilde j_{{M}'}$ are $2\pi-2\beta_2$.
 \end{itemize}      
 
Let us consider a special class of wormhole geometries that satisfy the requirements discussed below.  Suppose that the end points of the cuts  on the tree $j_{{N}},j_{{N}'},j_{{M}},j_{{M}'}$ are very close to  the end points  $\tilde j_{{N}}, \tilde j_{{N}'}, \tilde j_{{M}},\tilde j_{{M}'} $ respectively. Consequently,  the points $i_{{N}},i_{{N}'},i_{{M}},i_{{M}'}$ can also be made very  close to  the  points  $\tilde i_{{N}}, \tilde i_{{N}'}, \tilde i_{{M}},\tilde i_{{M}'} $ respectively. We also assume that after the Lorentzian continuation at late time the points $i_{A}$ and $j_A$ are very close to the branch point $q_A$ and  the points $i_{{A'}}$ and $ j_{A'}$ is very close to the branch point $b$ for $A=M,N$.   We shall show below that the least free energy wormhole geometry in this class has time independent free energy. The contributions to the saddle point approximation of the JT gravity partition function coming  from a conical singularity or corner depends on the value of the angle at the conical singularity or corner, the distance from the nearby conical singularity or corner and the value of the dilaton at that location. Therefore, the universal contribution to the free energy $Z\left(\mathbb{S}_{4n_1+4n_2+1};\tilde e_1, e_{1},\cdots,\tilde e_{2n_1+4n_2+4}, e_{2n_1+2n_2+4}\right)$  comes from the end points of the cuts  in $\mathbb{S}_{4n_1+4n_2+1}$. Moreover, these universal contributions are insensitive to the boundary conditions $e'$s and $\tilde e'$s. Therefore, we can identify them with the universal contribution to the free energy $Z\left({\mathcal{S}_{4n_1+4n_2+1}}\right)$
    \begin{align}\label{Zsa}
 Z_{universal}\left({\mathcal{S}_{4n_1+4n_2+1}}\right)\approx  e^{F_W} \end{align}
 where $F_W$ is the universal contributions to the free energy.  After the Lorentzian continuation, it is  given by 
   \begin{align}\label{Fg12}
&F_W \approx -\text{ln}~Z\left(\mathbb{S}_{4n_1+4n_2+1};\tilde e_1, e_{1},\cdots,\tilde e_{2n_1+4n_2+4}, e_{2n_1+2n_2+4}\right) \nonumber\\
& \approx \frac{1}{2} ca_1  \ln \left((w^+_{ i_{N'}}-w^+_{b})(w^+_{i_{M'}}-w^+_{b})\right)
\nonumber\\
&+  \frac{1}{2} ca_1\ln \left((w^-_{ i_{N'}}-w^-_{b})(w^-_{i_{M'}}-w^-_{b})\right)
\nonumber\\
& +  \frac{c}{2}a_2\ln \left((w^+_{j_N}-w^+_{q_N})(w^+_{j_{N'}}-w^+_{b})(w^+_{j_M}-w^+_{q_M})(w^+_{j_{M'}}-w^+_{b})\right)
\nonumber\\
& + \frac{c}{2}a_2 \ln \left((w^-_{j_N}-w^-_{q_N})(w^-_{j_{N'}}-w^-_{b})(w^-_{j_M}-w^-_{q_M})(w^-_{j_{M'}}-w^-_{b})\right)
\nonumber\\
&-\frac{a_3}{2 \pi G_N}\left(\Phi\left(w^+_{j_N}, w^-_{j_N}\right)+\Phi\left(w^+_{j_{N'}}, w^-_{j_{N'}}\right)+\Phi\left(w^+_{ j_M}, w^-_{j_M}\right)+\Phi\left(w^+_{j_{M'}}, w^-_{\tilde i_{M'}}\right)\right)\nonumber\\
&-\frac{a_4}{2 \pi G_N}\left(\Phi\left(w^+_{i_N}, w^-_{i_N}\right)+\Phi\left(w^+_{i_{N'}}, w^-_{i_{N'}}\right)+\Phi\left(w^+_{\tilde i_M}, w^-_{\tilde i_M}\right)+\Phi\left(w^+_{\tilde i_{M'}}, w^-_{\tilde i_{M'}}\right)\right)
\end{align}
  where 
  \begin{align*}
  a_1&=n_1\Theta(2\alpha_1-2\pi)+n_2\Theta(2\alpha_2-2\pi) \qquad a_2=\sum_{i=1}^2\left(\Theta(2\pi n_i)+\Theta(2\beta_i-2\pi)\right)\nonumber\\
     a_3&=\beta_1+\beta_2-2\pi \qquad a_4=n_1\alpha_1+n_2\alpha_2-(n_1+n_2)\pi.
  \end{align*}
   Minimising $F_W$ with respect to the coordinates of the end points of the cuts in the gravitational regions determines them, and completes the determination of the wormhole saddle. We shall denote this wormhole saddle as $\mathcal{S}_{4n_1+4n_2+1}^W$. It is possible to verify that the free energy of the wormhole saddle doesn't depend on the boundary time $u$. Therefore, the wormhole saddle dominates the trivial saddle at late times.

\section{Action of   \texorpdfstring{$\mathbb{U}_{bath}(t,u)$}{Lg}  in the semiclassical limit}\label{sect 5}

In this section we will study the nature of the action of $\mathbb{U}_{bath}(u,s)$ on a local operator before and after the Page time in the semiclassical limit.

   \subsection{Before Page time}
   Before Page time the dominant gravitational saddle is the trivial saddle.  This implies that before the Page time  density matrix $\bm{\rho}$ in the full quantum JT gravity coupled to conformal matter in the semiclassical limit can be replaced by the corresponding density matrix $\rho$ in the conformal field theory. Therefore, the action of $\mathbb{U}_{bath}(t,u)$ before  Page time in the semiclassical limit can be identified with the action of the conformal field theory  operator  
   \begin{equation}
   \label{UT}
   U_T(t,u)\equiv e^{-\text{i}tK_{{M_u^{bath}}'}}{e}^{\text{i}tK_{{M}_u^{bath}}}{e}^{\text{i}tK_{{{{N}_{u}^{bath}}'}}}{e}^{-\text{i}tK_{{N^{bath}_{u-a}}}}
   \end{equation}
    where $K_A$ is the modular Hamiltonians 
    $$K_{A}=-\text{ln}~\rho_{A} \qquad  A=M_{u}^{bath},{M_{u}^{bath}}',N_{u-a}^{bath},{N_{u}^{bath}}'.$$  In a two dimensional conformal field theory the modular Hamiltonian for an interval $A=(a,b)$ is given by the boost operator  \cite{Arias:2018tmw}  
\begin{align}
K_A=\int_a^b dx~ \beta(x) T(x).
\end{align}
where $$\beta(x)= \frac{2\pi}{\left(\frac{1}{x-a}-\frac{1}{x-b}\right)}$$ is the local inverse temperature multiplying the energy density  $$T(x)=\frac{1}{2}j(x)^2.$$ The current operator  satisfy the commutation relation
\begin{equation}
\label{currc}
\left[j(x),j(y) \right]=\text{i} \delta(x-y).
\end{equation}
The action of $U_T(t,u)$  on a primary operator in the conformal field theory shows that within the causal diamonds $\Diamond_{N_{u-a}^{bath}}$ and $\Diamond_{{N_{u}^{bath}}'}$ it acts as a translation operator. However, in the regions within $\Diamond_{M_u^{bath}}\cup\Diamond_{{M_u^{bath}}'}$ that are compliment to  $\Diamond_{N_u^{bath}}\cup\Diamond_{{N_{u}^{bath}}'}$ it acts as a boost operator. This implies that before the Page time in the semiclassical limit $\mathbb{U}_{bath}(t,u)$ can not transport a local operator in the bath to the black hole interior. Hence, as expected, this operator cannot reconstruct the black hole interior operators using the bath degrees of freedom before the Page time.

   \subsection{After  Page time}

In section \ref{wsaddle}, we determined the wormhole saddle that describe the expectation value of the local operator $\mathcal{O}$ that is acted by $\mathbb{U}_{bath}(t,u)$ in the semiclassical limit. In this subsection, we  will  identify the semiclassical description of  $\mathbb{U}_{bath}(t,u)$ after  Page time from the CFT path integral over the wormhole saddle. For this we shall express the CFT path integral as an expectation value of  $\mathcal{O}$ acted by an operator placed along the modified cuts in a $w$-plane for $n_1= it$ and $n_2= -it$. We will do this by interpreting the wormhole saddle as the result of the back reaction of the operator $\mathbb{U}_{bath}(t,u)$ placed in the $w$-plane in the $G_N\to 0$ limit. This can be conveniently done if  $\mathbb{U}_{bath}(t,u)$ in the semiclassical limit is an operator close to identity. Then the modified cut in which the operator is placed is not significantly different from the back-reacted geometry. As a result,  the modified cut in the $w$-plane  can be easily identified if $n_1$ and $n_2$ are assumed to be infinitesimal. \par

Consider the wormhole saddle ${\mathcal{S}}^W_{4n_1+4n_2+1}$ for infinitesimal values of $n_1=\epsilon_1$ and $n_2=\epsilon_2$. The associated deficit angles satisfying \eqref{alphabeta} can be approximated as
\begin{align}\label{approxalphabeta}
 \alpha_i=\left(1-\epsilon_i\right)\pi  \qquad \beta_i=\left(1-\epsilon_i+\epsilon_i^2\right)\pi \qquad\qquad i=1,2 .
 \end{align}
Since the wormhole saddle describing the action of infinitesimal reduced half-sided translation is obtained by setting $n_1=-n_2=it$, we assume that $$\epsilon_2=-\epsilon_1=\epsilon.$$ The free energy of this wormhole configuration is given by 
   \begin{align}\label{Fg13}
F_W& \approx  -\frac{1}{6} c\epsilon^2 \ln \left((w^+_{ i_{N'}}-w^+_{b})(w^+_{i_{M'}}-w^+_{b})\right)- \frac{1}{6} c\epsilon^2 \ln \left((w^-_{ i_{N'}}-w^-_{b})(w^-_{i_{M'}}-w^-_{b})\right)
\nonumber\\
& + \frac{1}{6} c\epsilon^2 \ln \left((w^+_{j_N}-w^+_{q_N})(w^+_{j_{N'}}-w^+_{b})(w^+_{j_M}-w^+_{q_M})(w^+_{j_{M'}}-w^+_{b})\right)
\nonumber\\
& + \frac{1}{6} c\epsilon^2 \ln \left((w^-_{j_N}-w^-_{q_N})(w^-_{j_{N'}}-w^-_{b})(w^-_{j_M}-w^-_{q_M})(w^-_{j_{M'}}-w^-_{b})\right)
\nonumber\\
&-\frac{\epsilon^2}{ G_N}\left(\Phi\left(w^+_{j_N}, w^-_{j_N}\right)+\Phi\left(w^+_{j_{N'}}, w^-_{j_{N'}}\right)+\Phi\left(w^+_{ j_M}, w^-_{j_M}\right)+\Phi\left(w^+_{j_{M'}}, w^-_{\tilde i_{M'}}\right)\right)\nonumber\\
&+\frac{\epsilon^2}{ G_N}\left(\Phi\left(w^+_{i_N}, w^-_{i_N}\right)+\Phi\left(w^+_{i_{N'}}, w^-_{i_{N'}}\right)+\Phi\left(w^+_{\tilde i_M}, w^-_{\tilde i_M}\right)+\Phi\left(w^+_{\tilde i_{M'}}, w^-_{\tilde i_{M'}}\right)\right).
\end{align}
 The locations of the end points of the cuts that minimises the free energy are given by
   \begin{align}\label{MN}
  &\left(w^+_{ i_M}, w^-_{ i_M}\right)= \left(w^+_{ j_M}, w^-_{  j_M}\right) =\left(-\frac{G_N\beta}{6\pi \phi_r}\frac{1}{w^-_{q_M}},-\frac{G_N\beta c}{6\pi \phi_r} \frac{1}{w^+_{q_M}}\right)\nonumber\\
         &\left(w^+_{ i_N}, w^-_{ i_N}\right)= \left(w^+_{  j_N}, w^-_{  j_N}\right) =\left(-\frac{G_N\beta}{6\pi \phi_r}\frac{1}{w^-_{q_N}},-\frac{G_N\beta c}{6\pi \phi_r} \frac{1}{w^+_{q_N}}\right)\nonumber\\
     &\left(w^+_{ i_{M'}}, w^-_{ i_{M'}}\right) = \left(w^+_{  j_{M'}}, w^-_{  j_{M'}}\right) = \left(-\frac{G_N\beta c}{6\pi \phi_r}\frac{1}{w^-_{b}},-\frac{G_N\beta c}{6\pi \phi_r} \frac{1}{w^+_{b}}\right)\nonumber\\
     &\left(w^+_{ i_{N'}}, w^-_{ i_{N'}}\right) =  \left(w_{  j_{N'}},\bar w_{  j_{N'}}\right) = \left(-\frac{G_N\beta c}{6\pi \phi_r}\frac{1}{w^-_{b}},-\frac{kG_N\beta c}{6\pi \phi_r} \frac{1}{w^+_{b}}\right).
   \end{align}
This determines the modified cuts in the $w$-plane. They are given by $${{\bm M}}_u^{bath}=\left(i_M,i_{M'}\right)\cup\left(q_M,i_0^R\right),$$  and $${{\bm N}}_{u-a}^{bath}=\left(i_N,i_{N'}\right)\cup\left(q_N,i_0^R\right),$$ as shown in figure \ref{fig:BHAdS2bathMNMod2}.  Therefore the semiclassical limit of  $\mathbb{U}_{bath}(t,u)$ after the Page time can be identified as  
   \begin{equation}\label{MredHST}
 \lim_{G_N\to 0} \mathbb{U}_{bath}(t,u)= \rho^{-\text{i}t}_{{{\bm M}_u^{bath}}} \rho^{\text{i}t}_{{{ M}'}_u^{bath}}\rho^{\text{i}t}_{{{\bm N}_{u-a}^{bath}}}  \rho^{-\text{i}t}_{{{ N}'}_{u}^{bath}} \qquad u>u_{Page}
   \end{equation}
   where, $\rho_{{{\bm M}_u}^{bath}}$ is the density matrix  associated with the region ${{\bm M}_u}^{bath}$ and  $\rho_{{{\bm N}_{u-a}^{bath}}}$ is the density matrix associated with the region ${{\bm N}_{u-a}^{bath}}$.


 \begin{figure}
\centering
\begin{tikzpicture}[scale=1.25]
    \draw [red,  thick] plot [smooth] coordinates {(-6,4.75) (-3.5,3.5) (-1,4.75)} ;
    \draw [thick,](1 mm, 10 pt) (-3.5,7)--(1.5,2.5) ;
        \draw [thick,](1 mm, 10 pt) (-3.5,-2)--(1.5,2.5) ;
      \draw  [red,  thick] plot[smooth]coordinates {(-6,0.25) (-3.5,1.5) (-1,0.25)} ;
     \draw [thick,](1 mm, 10 pt) (-3.5,7)--(-8.5,2.5)  ;
        \draw [thick,](1 mm, 10 pt) (-3.5,-2)--(-8.5,2.5) ; 
   \fill[fill=black!50!green] (-1,4.25)--(1.25,2.25)--(1.5,2.5)--(-.75,4.5);
    \fill[fill=red] (-4,2.8)--(-3.5, 2.35)--(-3,2.8)-- (-3.5,3.3);
     \fill[fill=yellow] (-4,2.8)--(-3.5, 2.35)--(-3.1,2.7)-- (-3.6,3.2);
  \fill[fill=blue!20] (-.75,4.025)--(1.25,2.275)--(1.5,2.5)--(-.5,4.275);
   \draw (-3,2.8) node  [circle,fill,inner sep=1pt]{};
    \draw (-3.1,2.7) node  [circle,fill,inner sep=1pt]{};
     \draw (-4,2.8)node  [circle,fill,inner sep=1pt]{};
      \draw (-3,2.7) node[below] {$j_{N}$};
      \draw (-3,2.8) node[right] {$j_{M}$};
     \draw (-4,2.9) node[left] {$j_{M'}$};
\draw (-1.3,4.1) node {$q_M$};
\draw (1.5,2.5) node  [circle,fill,inner sep=1pt]{};
\draw (1.8,2.55) node {$i_0^R$};
\draw (-8.5,2.5) node  [circle,fill,inner sep=1pt]{};
\draw (-8.8,2.55) node {$i_0^L$};
\draw (-5.6,4.3) node {$\Sigma_u$};
\draw (-5.4,3.5) node {$\Sigma_{u-a}$};
\draw (-1,4.25) node  [circle,fill,inner sep=1pt]{};
\draw (-3.5,2.5) node  [circle,fill,inner sep=1pt]{};
\draw (-.75,4.025) node[below] {$q_N$};
\draw (-.75,4.025) node  [circle,fill,inner sep=1pt]{};
  \draw [very thick](1 mm, 10 pt) (-6,4.75)--(-1,.25) ;
 \draw [very thick](1 mm, 10 pt) (-6,.25)--(-1,4.75);
 \draw[thick](1 mm, 10 pt) (-6,.25)--(-6,4.75);
  \draw[thick](1 mm, 10 pt) (-1,.25)--(-1,4.75);
       \draw [blue, very thick,dashed] plot [smooth] coordinates {(-8.5,2.5) (-6,4.25)(-4.05,2.8) (-3.5,2.5)(-3.05,2.8) (-1,4.25)(1.5,2.5)} ;
        \draw [violet, very thick,dashed] plot [smooth] coordinates {(-8.5,2.5) (-6.25,4.025)(-4.1,2.725) (-3.5,2.5)(-3,2.725) (-.75,4.025)(1.5,2.5)} ;
\end{tikzpicture}
\caption{ The wormhole saddle provides  modified semiclassical density matrices  $\rho_{{{\bm M}_u}^{bath}}$ and $\rho_{{{\bm N}_{u-a}^{bath}}}$. The domain of dependence of  the interval ${{\bm M}_u}^{bath}$ is  the union of the blue, green, red and the yellow shaded regions. The domain of dependence of the the interval ${{\bm N}_{u-a}^{bath}}$ is  the union of the blue and yellow shaded regions. \label{fig:BHAdS2bathMNMod2} } 
\end{figure}
   
     \subsection{Modular flow for a free chiral scalar in two intervals}\label{2interval}
    
  The action of the $\mathbb{U}_{bath}(t,u)$  on a local operator in the $G_N\to 0$ limit can be found if it is possible to compute the action of the reduced density matrices on them. Unfortunately, the reduced density matrices associated with multiple disconnected intervals  is know only for special CFTs. As argued in the previous subsections, the density matrices $\rho_{{{\bm M}_u}^{bath}}$ and $\rho_{{{\bm N}_u}^{bath}}$ that appear in  the semiclassical expression  $\mathbb{U}_{bath}(t,u)$ after the Page time are  associated with two disconnected intervals. Therefore, we will work with a special  CFT, namely a free massless scalar field in two dimension.  \par
   
   Consider the algebra $\mathcal{A}$ generated by the current $j(w^+)=\partial_{w^+}\psi\left(w^+\right)$ in a free massless scalar field  $\psi$ in the $w$-plane for a two interval region $A=A_1\cup A_2$ on the $w^+$ line. The interval $A_1$ is given by $a_1<w^+<b_1$, and the interval $A_2$ is given by $a_2<w^+<b_2$. An element of the operator  algebra $\mathcal{A}$ is given by 
   \begin{equation}
   \label{osmear}
   \mathcal{O}^-=\int_{A} dw^+  \gamma(w^+) j\left(w^+\right)
   \end{equation}
    where $ \gamma(w^+)$ is an arbitrary distribution. We shall  describe the modular evolution of this element. \par
   
    The modular Hamiltonian that generate the modular flow contains both a local part and a non-local part \cite{Arias:2018tmw} 
\begin{align}\label{2intK}
K_{A}=-\text{ln}~\rho_A=K_{A}^{\text{local}}+K_{A}^{\text{nolocal}}
\end{align}  
where $\rho_A$ is reduced density matrix associated with the interval $A$. The local part of the modular Hamiltonian is  
\begin{align}\label{Loc}
K_{A}^{\text{local}}=\int_A dx~ \beta(x) T(x),
\end{align}
where $$\beta(x)= \frac{2\pi}{\sum_i\left(\frac{1}{x-a_i}-\frac{1}{x-b_i}\right)}$$ is the local inverse temperature multiplying the energy density  $$T(x)=\frac{1}{2}j(x)^2.$$ The non local part of has the following structure
\begin{align}\label{NonL}
K_{A}^{\text{nolocal}}&=\int_{A_1\times A_1 }dxdy~ j(x)N(x,y)j(y)-\int_{A_1\times A_2 }dxdy~ j(x)N(x,\bar y)j(y)\nonumber\\
&-\int_{A_2\times A_1 }dxdy~ j(x)N(\bar x,y)j(y)+\int_{A_2\times A_2 }dxdy~ j(x)N(\bar x,\bar y)j(y). 
\end{align}
where the map $$\bar x=\frac{a_1a_2(x-b_1-b_2)-b_1b_2(x-a_1-a_2)}{x(a_1+a_2-b_1-b_2)+(b_1b_2-a_1a_2)}$$ interchanges the two intervals. 
The functions $N(x,y)$ is an integrable function do not identically vanish in any open set of $A\times A$, with at most $\text{ln}|x-y|$ singularity for $x\sim y$.  Then the smearing function of  
\begin{equation}
\label{oflowed}
\mathcal{O}^+(t)=\rho_A^{-\text{i}t}\mathcal{O}^+\rho_A^{\text{i}t}
\end{equation}
 satisfies the following linear equation
\begin{align}
\partial_{t}\gamma(w^+,t)=-\beta(w^+)\partial_{w^+}\gamma(w^+,t)-\int_{A} d\tilde w^+ \tilde N(w^+,\tilde w^+)\partial_{\tilde w^+}\gamma(\tilde w^+,t)
\end{align}
where $\tilde N(w^+,\tilde w^+)$ is a function that do not identically vanish in any open set of $A\times A$.  It is a function that built using $N(w^+,\tilde w^+)$. The explicit form of this function can be found in \cite{Arias:2018tmw}. This equation suggests that if we start with a smearing function $\gamma(w^+,0)$ localised in one of the interval $A'_i\subset A_i$ at a finite distance from its boundary, then $\gamma(w^+,t)$ will spread everywhere in $A$. If $\gamma(w^+,0)$ localised  near a boundary of $A'$,  then the  spreading will  not happen due to the fact that the local inverse temperature $\omega$ near the boundary of $A'$ is infinitesimal. \par

      \subsection{Action of  \texorpdfstring{$\mathbb{U}_{bath}(t,u)$}{Lg}  after  Page time}

   Assume that the matter conformal field theory in the $w$-plane is obtained by combining the two chiral theories as described in the previous subsection.   For this choice of matter theory, we can compute the action of $\mathbb{U}_{bath}(t,u)$ on a chiral operator $\mathcal{O}^+$ in the causal diamond $\Diamond_{{M}^{bath}_u}$  after the Page time in the semiclassical limit
   \begin{align}
{\mathcal{O}}_t^+&=\lim_{G_N\to 0}{\mathbb{U}_{bath}(t,u)}^{\dagger}{\mathcal{O}}^+\mathbb{U}_{bath}(t,u)\nonumber\\
&=  \rho^{\text{i}t}_{{{ N}'}_u^{bath}}\rho^{-\text{i}t}_{{{\bm N}_u^{bath}}} \rho^{-\text{i}t}_{{{ M}'}_u^{bath}}  \rho^{\text{i}t}_{{{\bm M}_u^{bath}}} {\mathcal{O}}^+\rho^{-\text{i}t}_{{{\bm M}_u^{bath}}} \rho^{\text{i}t}_{{{ M}'}_u^{bath}}\rho^{\text{i}t}_{{{\bm N}_u^{bath}}}  \rho^{-\text{i}t}_{{{ N}'}_u^{bath}} \nonumber\\
&=\rho^{-\text{i}t}_{{{\bm N}}_u^{bath}}\rho^{\text{i}t}_{{{\bm M}}_u^{bath}}  {\mathcal{O}}^+\rho^{-\text{i}t}_{{{\bm M}}_u^{bath}}  \rho^{\text{i}t}_{{{\bm N}}_u^{bath}}
\end{align}
  Following the discussion in section \ref{2interval}, it is straightforward verify  that $\mathbb{U}_{bath}(t,u)$ spreads the operator $\mathcal{O}^+$   to the domain of dependence of the interval $\left(j_M,j_{M'}\right)$, the island region, which has nonzero intersection with the black hole  interior, see figure \ref{fig:BHAdS2bathMNMod2}.  Therefore, we conclude that $\mathbb{U}_{bath}(t,u)$ reconstructs the operators inside the black hole interior after the Page time. 



\section{Conclusion}\label{sect 6}

In this paper, we studied a special operator $\mathbb{U}_{bath}$ that has nontrivial action only on the bath degrees of freedom. We showed via the gravitational Euclidean path integral analysis that though before the Page time  $\mathbb{U}_{bath}$ in the semiclassical limit does not transport the operators in the bath to the black hole interior, after the Page time it takes them to the black hole interior. This demonstrates that after the Page time the operators in the black hole interior can be reconstructed using the bath degrees of freedom which agrees with the assertions of the island paradigm.  It will be interesting to investigate how this procedure is realised if the matter CFT has a dual three dimensional gravitational description. Also, it might be interesting to study the relevance of the operator $\mathbb{U}_{bath}$ from the perspective of algebraic quantum field theory.

\section*{Acknowledgements}
We are grateful to Sujay Ashok,  Alok Laddha, Abhijit Gadde, Sunil Mukhi,  Onkar Parrikar and Sandip Trivedi for the  discussions and comments. We also thank Horacio Casini for a helpful clarification. 


\appendix



\end{document}